\documentclass[pre,showpacs,amsmath,onecolumn]{revtex4}  
\usepackage{graphicx}
\usepackage{amssymb}
\usepackage{bm}
\usepackage{amsmath}
\usepackage{amsfonts}
\usepackage{subfig}

\begin{document}
\title{Macroscopic model of
self-propelled bacteria swarming with regular reversals}

\author{ Richard Gejji$^{1,2}$, Pavel M. Lushnikov$^{3}$,   and Mark
 Alber$^{1*}$
}

\affiliation{$^1$ Department of Applied and Computational
Mathematics and Statistics, University
of Notre Dame, Notre Dame, IN 46656, USA \\
$^2$ Mathematical Biosciences Institute, Ohio State University, 1735 Neil Avenue, Columbus, OH 43210\\
 $^3$ Department of Mathematics and Statistics,
 University of New Mexico, Albuquerque, NM 87131, USA}

\date{\today}


\begin{abstract}
Periodic reversals of the direction of motion in systems of
self-propelled rod shaped bacteria enable them to effectively
resolve traffic jams formed during swarming and maximize their
swarming rate. In this paper, a connection is found between a
microscopic one dimensional cell-based stochastic model of reversing
non-overlapping bacteria and a macroscopic  non-linear diffusion
equation describing dynamics of the cellular density.
Boltzmann-Matano analysis is used to determine the nonlinear
diffusion equation corresponding to the specific reversal frequency.
Macroscopically (ensemble-vise) averaged stochastic dynamics is
shown to be in a very good agreement with the numerical solutions of
the nonlinear diffusion equation. Critical density $p_0$ is obtained
such that nonlinear diffusion
 is dominated by the collisions between cells for the densities
$p>p_0$. An analytical approximation of the pairwise collision time
and semi-analytical fit for the total jam time per reversal period
are also obtained. It is shown that cell populations with high
reversal frequencies are able to spread out effectively at high
densities. If the cells rarely reverse then they are able to spread
out at lower densities but are less efficient at spreading  out at at higher densities.
\end{abstract}

\pacs{ 87.18.Ed, 05.40.-a, 05.65.+b, 87.18.Hf,  87.10.Ed; 87.10.Rt}

\maketitle
\section{Introduction}

Many bacteria including species found in diverse soil and water
environments are able to spread rapidly over surfaces by the process
of swarming which is the collective motion of  many thousands of
cells. The bacteria capable of swarming span the gamut of utilit
and range from innocuous carbon-cycle organisms to harmful
pathogens. Swarming can be achieved by directional movement from
pulling with type IV pili and either propulsion from rotating
flagella or pushing from secretion of slime \cite{Kaiser07}. In
certain cases, these mechanisms work together and allow the cells to
swarm at a rate faster than each individual type of motility
\cite{Wu07,Kaiser83}.

\begin{figure}
  \begin{center}
  \subfloat[]{\label{myxoCluster}\includegraphics[scale=0.25]{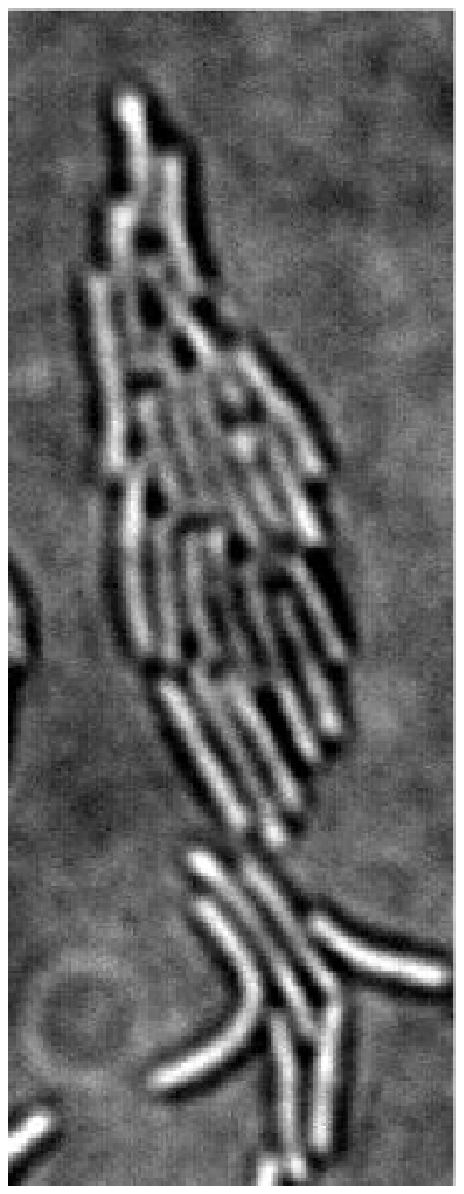}}
  \quad  \quad
 \subfloat[]{\label{myxoSwarm}\includegraphics[scale=0.25]{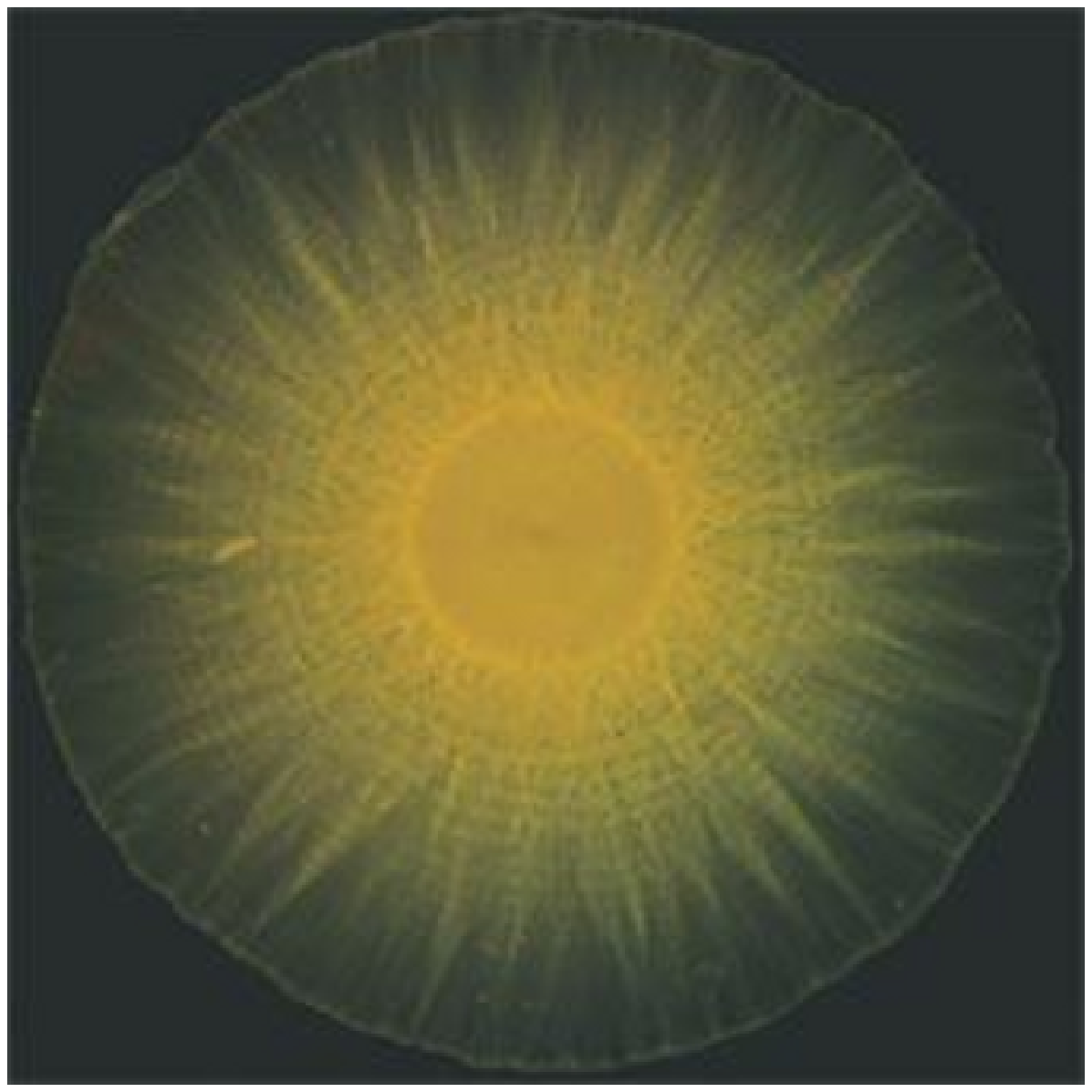}}
 \quad \quad
   \subfloat[]{\label{cellExpansion}\includegraphics[scale=0.27]{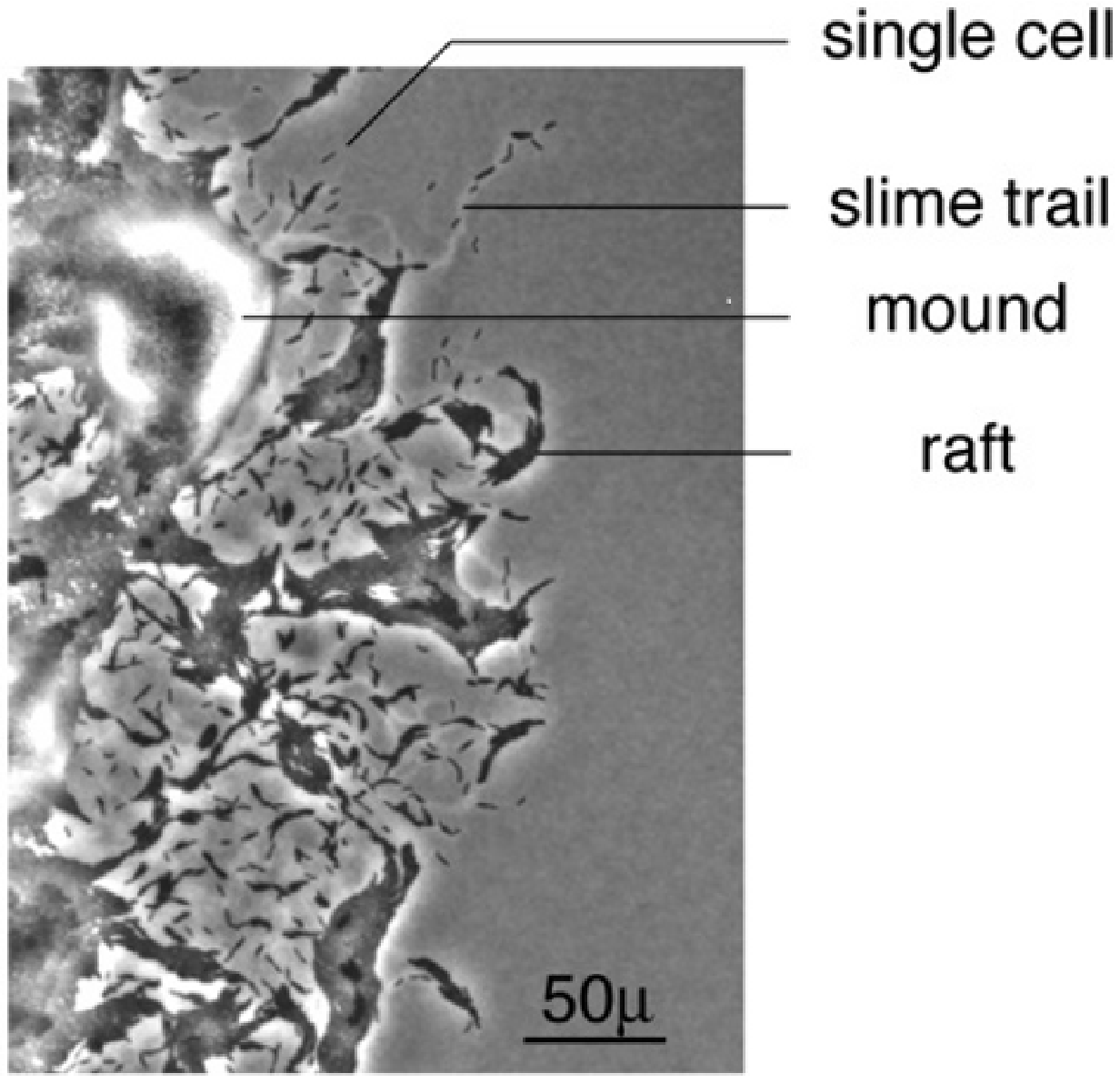}}
\end{center}
\caption{a) Myxobacteria aligned together into a cluster. Picture
obtained in Dr. Shrout's lab by Cameron Harvey. b) Swarm of \emph{M.
Xanthus}, picture taken by Lotte Jelsbak. The edge of the swarm
displays a single layer of cells that are spreading outwards away
from the cell center \cite{Kaiser07} c) Distribution of cells at the
swarm edge \cite{Wu09}. The multicellular structures, slime trails,
mounds, and rafts are labeled. The swarm is expanding in the radial
direction, which is to the right of the image.}
\label{experimentImages}
\end{figure}

For example, {\it Myxococcus xanthus}, ubiquitous bacteria found in
soil, are very efficient swarmers. These bacteria have elongated
rod-type shapes (about $7 \mu m$ in length and $0.5 \mu m$ in width)
and they move by gliding over a substrate in the direction of their
longer axis \cite{Kaiser07,Wu07,Yu07,Wu09}.  They align and travel
together in the same direction (see Figure \ref{experimentImages}a)
as well as reverse direction of their motion about every eight minutes \cite {Kaiser07,Wu09,Mignot05}. Mutant species of
Myxobacteria that are unable to reverse are also unable to swarm
\cite{Myxobacteria,Wu09}.

After agar plate is inoculated in the center with {\it M. xanthus},
they start growing and moving, and the swarm expands. 90\% of the
expansion is caused by cell movement and only 10\% by growth
\cite{Kaiser83}. It has been shown that a reversal period of $8.8
min$ maximizes the expansion rate for a given average cell velocity
of $4\mu m/min$ \cite{Wu09}. Such motion is limited by new cells
moving out from the center. Therefore, a cell in many cases can not
move full 8 minutes in the direction towards the center. When
encountering a cell moving in opposite direction cell stops and
waits till it is time to start moving again away from the center.
The swarm grows symmetrically in all directions (see Figure
\ref{experimentImages}b). The symmetry dictates that there is a net
movement only in radial directions.

The expansion rate of a wild type {\it M. xanthus} (A+S+) swarm
(which moves using both pili VI and slime engine) is $\sim 1.4\mu
m/min$ while the average velocity of individual myxobacteria is
$\sim 4 \mu m/min$ \cite{Wu09}. Because cells reverse periodically,
it is possible to say that $80\%$ of their energy is used for
swarming.  Also, large aspect ratio of cells and their ability to
bend promote bacterial alignment which also increases swarming.
Velocities of mutants (A-S+) and (A+S+) are equal to the velocity of
the wild type bacteria (A+S+).

Recent two-dimensional (2D) off-lattice microscopic stochastic model
(MSM) described in \cite{Wu07}, has been able to predict optimal
reversal rates for specific choices of bacterial velocities and aspect ratios leading to the maximal
swarming rates of the colony, which were confirmed in experiments
\cite{Wu09}. It has been also shown in \cite{Wu09} that such choice
of the optimal reversal rate allows cells to align better and
resolve traffic jams resulting in the maximal order of alignment.
The model takes into account cell shape and direction of motion of
each Myxobacteria in the colony determined by the two motility
mechanisms: pili VI and slime production. Such detailed cell
description is computationally intense and makes any analytical
description difficult.

Experimental observations suggest that the capacity to swarm depends
less on the motility engine employed by individual cells and instead
on the behavioral algorithms that enhance the flow of densely packed
cells \cite{Kaiser07,Wu09}. Because of that we focus in this paper
on the study of self-propelled motion of rod shaped bacteria without
specifying motility engines. We do not incorporate cell division,
the directional effects of slime, or the social motility governed by
pili into the model. Instead we use a basic model for studying
mechanism of swarming caused by the cell-cell collisions (jams) and
regular cell reversals, in a small part of the colony near the edge
of the swarm where motion of bacteria is nearly one-dimensional in
the radial direction (see Fig. \ref{experimentImages}).  We assume
that cells cannot climb onto each other and that no more than one
cell could be positioned at any moment in time at specific location
in space (by using excluded volume constraint). Cells are modeled as
self-propelled bending rods which glide on slime on the substrate
with the same constant velocity. They reverse direction of their
motion periodically in time which serves as a mechanism for
diffusion of spatial positions of cells.

The main result of the paper is an establishment of  a connection
between one-dimensional (1D) microscopic and macroscopic models and
and their parameters, describing swarming of bacteria reversing at
different frequencies. Microscopic 1D discrete stochastic model is a
1D simplified version of the full 2D model \cite{Wu07}. 1D
macroscopic model is a nonlinear diffusion equation for cellular
density describing dynamics of self-propelled non-overlapping rods
with regular reversals. Combination of a stochastic discrete
model and its continuous limit in the form of a nonlinear diffusion
equation constitutes a multi-scale modeling environment which allows
one to zoom in and study individual bacteria and then zoom out and
investigate emerging behavior of a large number of bacteria in a
swarming colony.

Although, only few continuous models based on biological cell
behavior exist which take volume of cells into account and prevent
cells from overlapping, such models are more biologically relevant
and can provide novel insights. Recently continuous limit models
describing dynamics of cellular density were derived from the
microscopic motion of randomly moving cells exhibiting volume
exclusion and chemotaxis \cite{Alber06,Alber07,Lushnikov08}. In
particular,  the Ref. \cite{Lushnikov08} introduces a nonlinear
diffusion equation model with chemotactic term for describing amoeba
aggregation without blow up of solution in finite time
\cite{Alber09}. This is in contrast with the standard but
biologically less realistic Keller-Segel equation (sometimes also
called Patlak-Keller-Segel equation) with constant diffusion
coefficient \cite{Patlak1953,KeSe1970} which neglects the size of
bacteria resulting in solution (bacterial density) having a blow up
(collapse) in finite time
 \cite{BrCoKaScVe1999,LushnikovPhysLettA2010,DejakLushnikovOvchinnikovSigalPhysDSubmitted2010}.

Another  1D continuous limit equation was recently derived from a
model of cells that interact using Hooke's Law \cite{Murray09}. This
equation also displays nonlinear fast diffusion, and looks similar
to the porous medium equation but with a negative exponent. This
model agrees well with the discrete system from which it is derived
and it is capable of effectively making biological predictions for
cells that can be modeled as stiff springs.

The paper is organized as follows. In Section
\ref{section:MicroscopicStochasticModelcellmotion} a Microscopic
Stochastic Model (MSM) of cellular dynamics is introduced which
describes 1D motion of sell-propelled rods with periodic in time
reversals of the direction of their motion. In Section
\ref{section:MSMSimulations} general settings for MSM simulations
are presented  and results of multiple 1D dynamics MSM simulations
with initially localized distributions of bacterial colonies are
described.
In Section \ref{section:elementarycollisions} elementary
laws of collisions (jams) between cells are derived and equilibrium
motion of cells is determined  in the limit of zero noise of the reversal period. Without interactions (in a vanishing  cell density
limit) each cell experiences almost periodic motion in space and time. Without a  noise in the reversal time that motion would be strictly periodic.
However, the experimentally observed \cite{Welch01} small noise in the reversal time results in the random walk of the average position   (averaged over time
period $2T$) of the center of mass of each cell at time scales above $2T$. Here $T$ is the average reversal time for each cell. For finite cellular densities we introduce different
types of collisions (jams) between cells including pairwise jam and
cluster jam. We also find the critical density $p_0$ below which
cellular diffusion is dominated by the diffusion (random walk) of individual cells while above $p_0$ the diffusion is dominated by the collisions between cells.
In Section \ref{section:multiplecollisions}
multiple collisions and cell clustering for large cellular densities
are studied. In Section
\ref{section:MacroscopicNonlinearDiffusionModelvsMSMsimulations}  a
nonlinear diffusion equation of the general form
\begin{eqnarray}\label{nonlineardiffusion1}
\partial _t p=\partial_{x} \Big [D(p)\partial_{x} p\Big ],
\end{eqnarray}
is introduced, where $p(x)$ is a local cell density (measured in units of volume
fraction, i.e. the ratio of volume occupied by cells to the total
volume of space), $x$ is the spatial coordinate and $D(p)$ is the
nonlinear diffusion coefficient determined by using Boltzmann-Matano
(BM) analysis \cite{BoltzMat} of ensemble averaged MSM simulations
of cells moving with different reversal frequencies. The equation
(\ref{nonlineardiffusion1}) gives microscopically averaged
dynamics of cellular density vs. microscopic description of MSM
model. We compare the dynamics of cellular density from MSM
simulations with the numerical solutions of the equation
(\ref{nonlineardiffusion1}) for different reversal frequencies and
find very good agreement between these two types of simulations for
$p>p_0$. This confirms that the dynamics of
cellular density is indeed of a nonlinear diffusion type
(\ref{nonlineardiffusion1}). In Section
\ref{section:Analyticalapproximation} an analytical approximation
for pairwise collision time and semi-analytical fit for the total
jam time per reversal period are described. In Section
\ref{section:ConclusionDiscussion} main results of the paper and
future directions are discussed. In Appendix A Boltzmann-Matano
analysis is reviewed. Appendix B provides additional testing of the
accuracy of Boltzmann-Matano approach for the cellular distributions
of finite size.

\section{Microscopic Stochastic Model of Bacterial Motion}
\label{section:MicroscopicStochasticModelcellmotion}

In this section we introduce a  computational discrete microscopic
stochastic model (MSM) of cellular dynamics describing motion of
self-propelled rods on a 1D lattice with periodic in time reversals
of the direction of their motion.

We simulate constant number of cells of length $L$ that move back
(left) and forth (right) in a spatial domain along the coordinate
$x$ with periodic boundary conditions  and velocity $v$. We assume that each cell reverses the direction of its motion in average  time $T$ after previous reversal.
The reversal period experiences fluctuations with the variance $\Delta T_0^2$ which are sharply peaked near $T$, i.e. $\Delta T_0/T\ll 1$ in accordance with the observed in experiments \cite{Welch01}.
Assume that the positive integer $n$ corresponds to the $n$th reversal of the given cell.
We chose the probability distribution for the stochastic realizations of the reversal time $T_n$ for $n$th reversal to be defined through the Poisson distribution
\begin{equation}\label{poisson1}
f(k)=\frac{\lambda^k e^{-\lambda}}{k!}, \quad k=0,1,2,\ldots
\end{equation}
as follows
\begin{equation}\label{Trandom}
T_n=k\Delta T_1,
\end{equation}
where $\lambda=T^2/\Delta T_0^2$ and  $\Delta T_1=\Delta T_0^2/T$. Because the statistical averages for (\ref{poisson1}) are $\langle f(k)\rangle=\lambda$ and $\langle (f(k)-\langle f(k)\rangle)^2\rangle =\lambda$
we obtain that $\langle T_n\rangle=T$ and $\langle (T_n-\langle T_n\rangle)^2\rangle =\Delta T_0^2$.  With that definition the stochastic realizations of $T_n$ can take only discrete
values $0,\Delta T_1,2\Delta T_1,\ldots$.
But because we assume $\Delta T_0/T\ll 1$ we conclude that this is a good approximation of the Myxobacteria reversals with continues set of values of the reversal time.

To quantify the difference between reversal times of neighboring bacteria we introduce the reversal phase for each bacteria defined as follows. We define time periods $(0,2T)$, $(2T,4T)$,  $(4T,6T)$ etc.
Inside each of these intervals each cell (bacteria) has an
assigned 
 reversal phase, $\phi$, between $0$ and
$2T$ corresponding to the time when cell reverses from moving to the
right to the motion to the left (i.e. cell reverses from
right-directed motion to the left-directed at times $t=\phi, \,
2T+\phi, \, 4T +\phi, \ldots$). Reversal in opposite direction
occurs at at times $t=\phi, \, T+\phi, \, 3T+\phi, \ldots$. The initial phase
of each cell is chosen at random.
Here the phase $\phi$ is not constant but changes after each reversal because of fluctuations of $T_n$. The condition $\Delta T_0/T\ll 1$ ensure that the change of $\phi$ at the time scale $T$ is small.
Respectively, at much larger time scale $t\gg T$ the phase $\phi$ experiences the random walk. These random walks are independent for for each of $N$ cells in the system.

\begin{figure}[thp]
\begin{center}
  \subfloat[]{\label{cellDiag}\includegraphics[width=160mm]{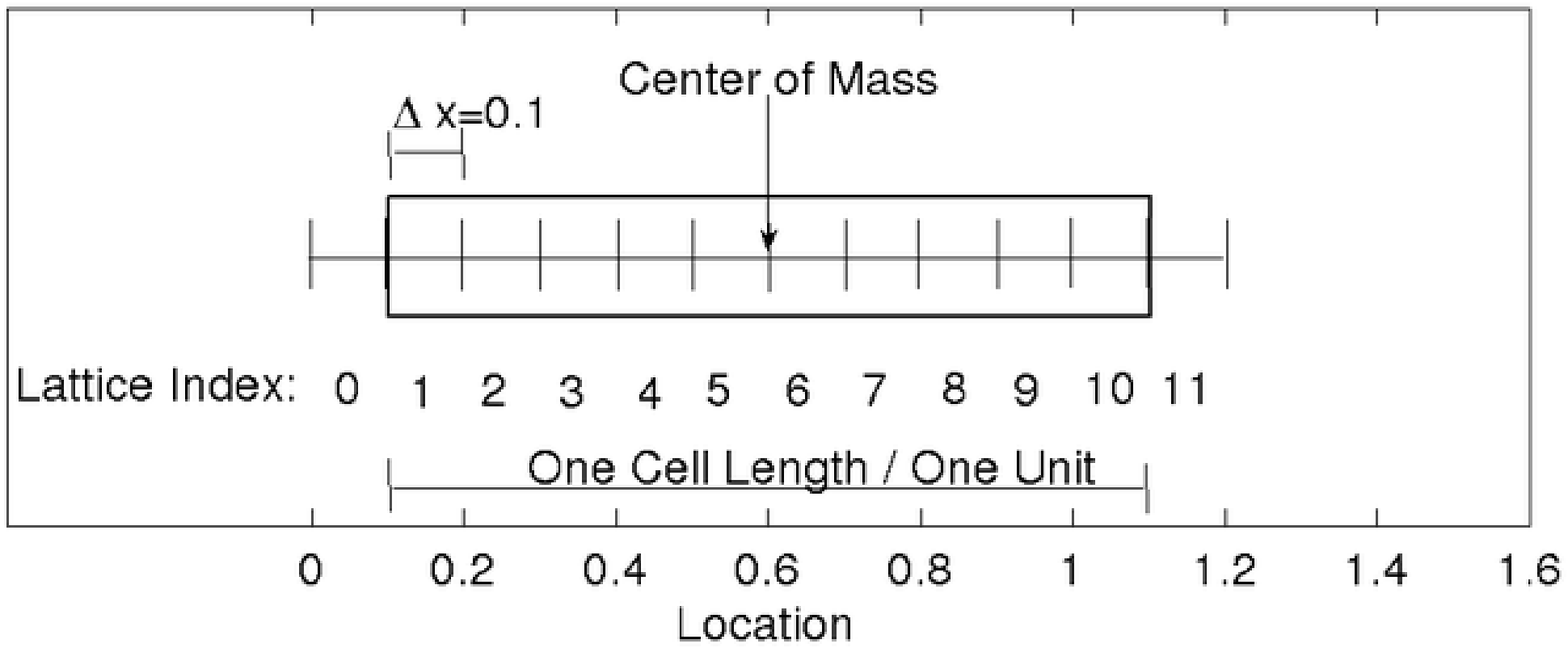}}\\
  \subfloat[]{\label{seqStepsA}\includegraphics[width=80mm]{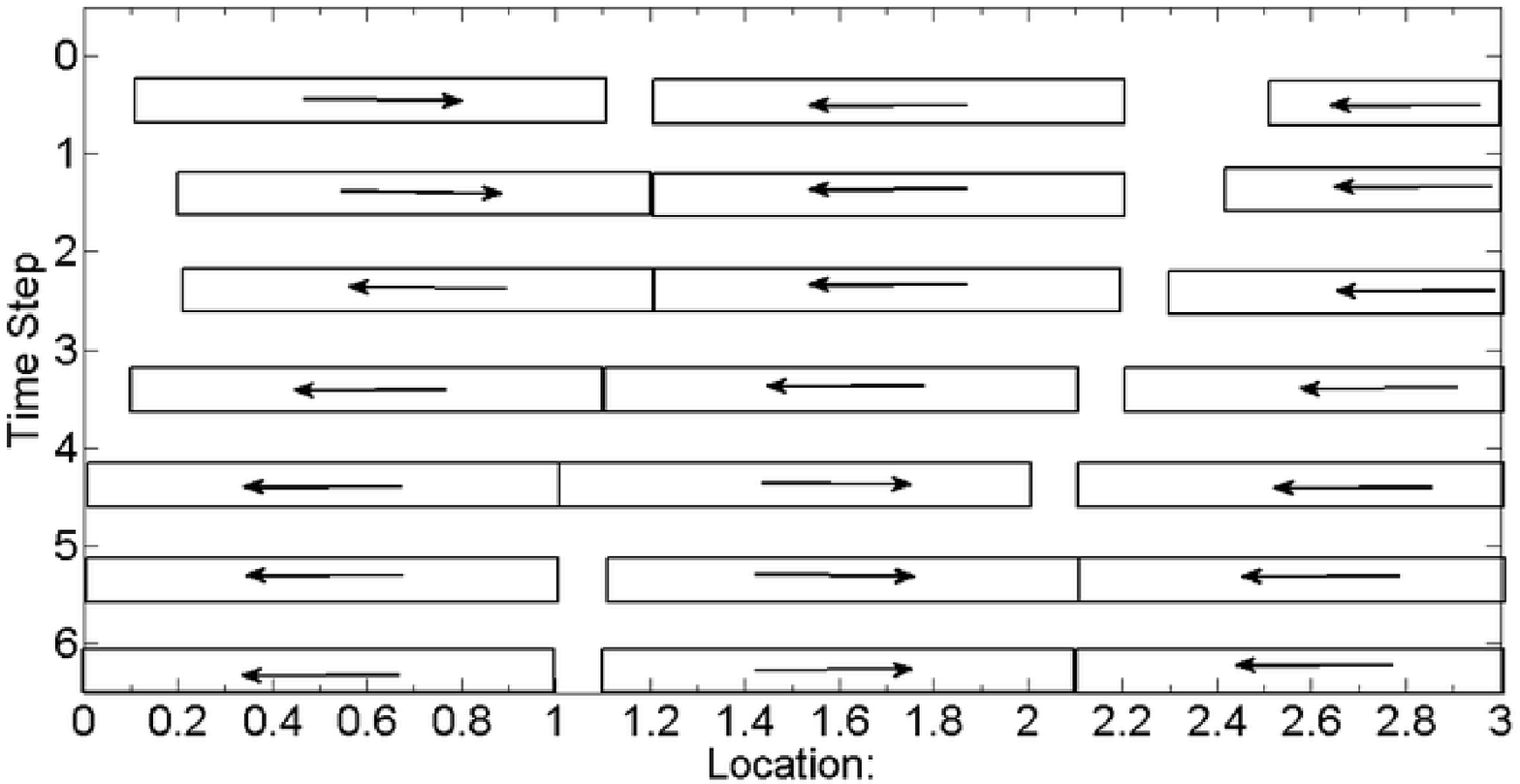}}
  \quad \quad
  \subfloat[]{\label{seqStepsB}\includegraphics[width=80mm]{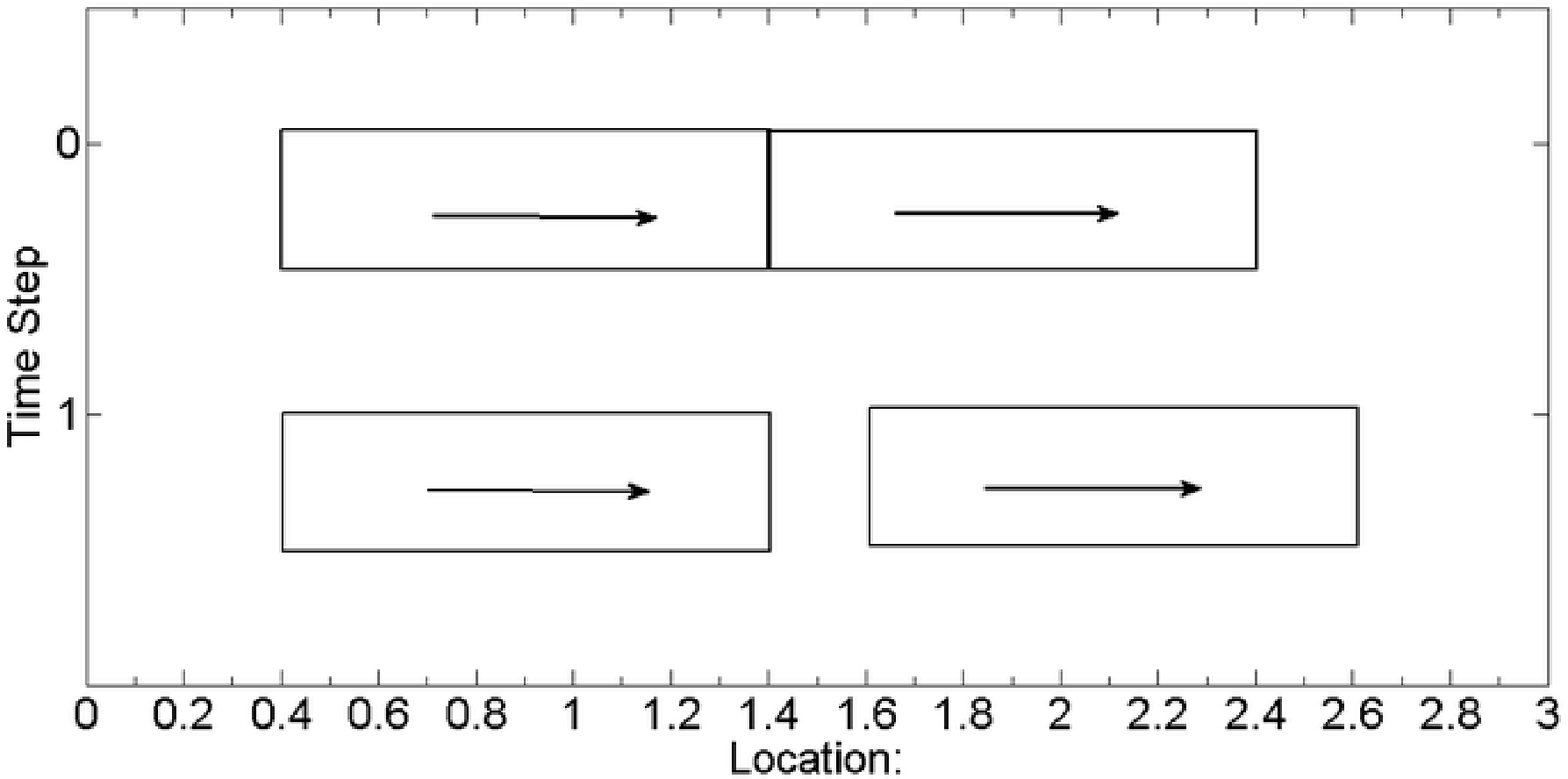}}
  \end{center}
\caption{a) Diagram of a single cell with marks designating absolute
location versus lattice index.  b) Sample movement in a single time
step where one cell is chosen twice and the other is not chosen at
all. c) Sample movement of three cells where each cell is chosen
exactly once.} \label{cellSketch}
\end{figure}

 In dimensionless units we assume
that $L=1$ and $v=1$. Unless otherwise specified we choose $T=8$.  Each cell is represented  by a finite number of lattice sites on a 1D grid. In a typical simulation, each cell
includes 10 lattice sites, i.e.  distance between neighboring
lattice sites is $\triangle x =0.1$ (see Figure \ref{cellSketch} a).
Time step in dimensionless units is $1/\triangle x$ to keep the velocity $v=1$. However, we also ran multiple simulations with smaller values of $\triangle x$
 to make sure that increasing number of lattice
sites per cell (but keeping $L=1$ and decreasing time step to keep
$v=1$) does not significantly change our results .
In other words, we look at the continuous limit as $\triangle x\to 0.$

The following three dimensionless parameters completely determine the
dynamics of cells in continuous limit $\triangle x\to 0.$  The first parameter is $vT/L,$ which is the ratio of the average
distance traveled by cells between reversals and the cell length.
That parameter is  $vT/L=8$ for the typical value $T=8$.
The second dimensionless parameter is the local cellular density
$p(x)$ measured in units of volume fraction $p$, i.e. the
ratio of volume occupied by cells to the total volume of space (in
1D, volume is simply the length). The third parameter is $\Delta T_0/T$, i.e. relative size of the reversal time fluctuations.

For example, we can choose the velocity, the reversal period, the fluctuation of the reversal period and the
cellular length as $v_{dim}=10\mu m/min$, $T_{dim}=8$min, $\Delta T_{0,dim}=0.9$min  and
$L_{dim}=10\mu m$, respectively, in dimensional units. This yields
$v_{dim} T_{dim}/L_{dim}=8$ similar to the typical dimensionless
values chosen above.  This choice is consistent with cell lengths
and reversal period used in previous computational models
\cite{Wu07,Wu09} and observed in experiments \cite{Myxobacteria}.
Below  unless otherwise specified we choose  $\Delta T_1= 0.1$. For $T=8$ it corresponds to   $\Delta T_0\simeq 0.9$.
Experiments with Myxobacteria typically show only small fluctuations
of the reversal period $T$ so that probability distribution function
is sharply peaked near average reversal period $T$ \cite{Welch01}. Our typical choice  $\Delta T_0\simeq 0.9$ reflects that smallness of fluctuations.
We also checked that if  there is no noise added to the reversal period, the simulations fails to match the nonlinear diffusion equation (\ref{nonlineardiffusion1}).
 This suggests that noise (although small) in  the reversal period  of bacteria contributes to their  macroscopic behavior by allowing them to behave diffusively.

At each time step, the model determines cell movement based on the
occupancy of the next lattice site in the direction the cell is
moving (direction is determined by $\phi$ at the given time $t$). If
this location is free, the cell is moved 1 lattice site in that
direction (keeping constant length $L=1$). If the location is not
free, the cell does not move.

The model itself is stochastic. During each time step, a sequence of
$N$ randomly chosen cells attempt to move one at a time, where $N$
is the total number of cells. It is possible that the same cell may
move more than once during a time step, and as a result some other
cells may not move at all. Also, note that random selection of cells
may create gaps between cells that are following each other (see
Figures \ref{cellSketch}b and \ref{cellSketch}c for examples of
possible cellular movement).
Creation of such gaps is equivalent to
the extra diffusion each cell experiences in addition to the
directed motion with the speed $v$. That diffusion is a pure
artefact of the finite width $\triangle x $ of each lattice site
which vanishes as $\triangle x \to 0.$ We checked in simulations
that reduction of $\triangle x$ from 0.1  to 0.001 gives only small
changes in the cellular density dynamics  (see Section \ref{section:MacroscopicNonlinearDiffusionModelvsMSMsimulationsSubsectionb} for more discussion on that).
However, the collision time between cells is more sensitive to the
value of $\triangle x$ so for these type of simulations in Section \ref{section:Analyticalapproximation} we also used grid spacing down to $\triangle x=0.001$.
Generally, in all quantities plotted in all Figures of the paper we
choose $\triangle x=0.1$ unless we explicitly specify a different
value of  $\triangle x$.

Unless otherwise specified, the simulations are run on a
one-dimensional lattice domain of length $4,000$ centered at $x=0$
starting with an initial top-hat distribution of cells of width
$1,000$ also centered at $x=0$ (i.e. density of cells is
approximately constant $p\equiv p_{max}$ for $-500<x<500$  and zero
everywhere else). See curve for $t=0$ in Figure
\ref{8reversalperiod}a as example of a top-hat boundary condition.
Because the domain is symmetric between $x$ and $-x$ it replicates a
no-flux boundary condition at $x=0$ after averaging over the
statistical ensemble of simulations. Generally, we choose  lattice
domain of length large enough to have no influence of the periodic
boundaries (i.e. to have zero cellular density at both right and
left boundaries).

\section{MSM  Simulations}
\label{section:MSMSimulations}

Swarming of bacterial colony similar to the one shown in Figure
\ref{experimentImages}a can be analyzed by averaging over angles,
i.e. as an average over dynamics of many nearly 1D (in the radial
direction) distributions of Myxobacteria. In what follows we assume
that motion along radius is a dominant one while rotation is only a
correction which we neglect here.

\begin{figure}[thp]
\begin{center}
 \subfloat[]{\label{cellDensOverTime}\includegraphics[width=80mm]{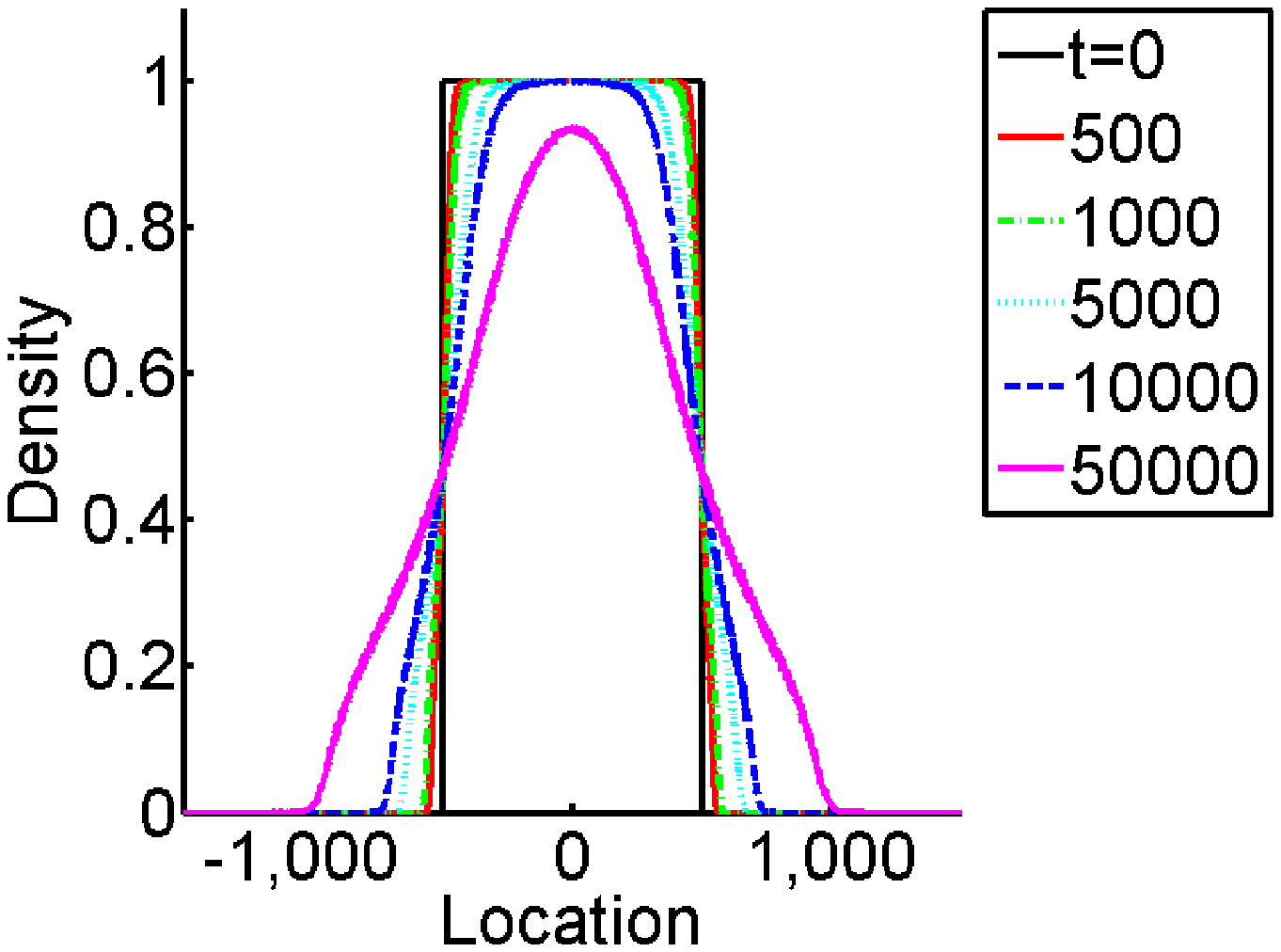}}\quad
 \subfloat[]{\label{expecClustOverTime}\includegraphics[width=80mm]{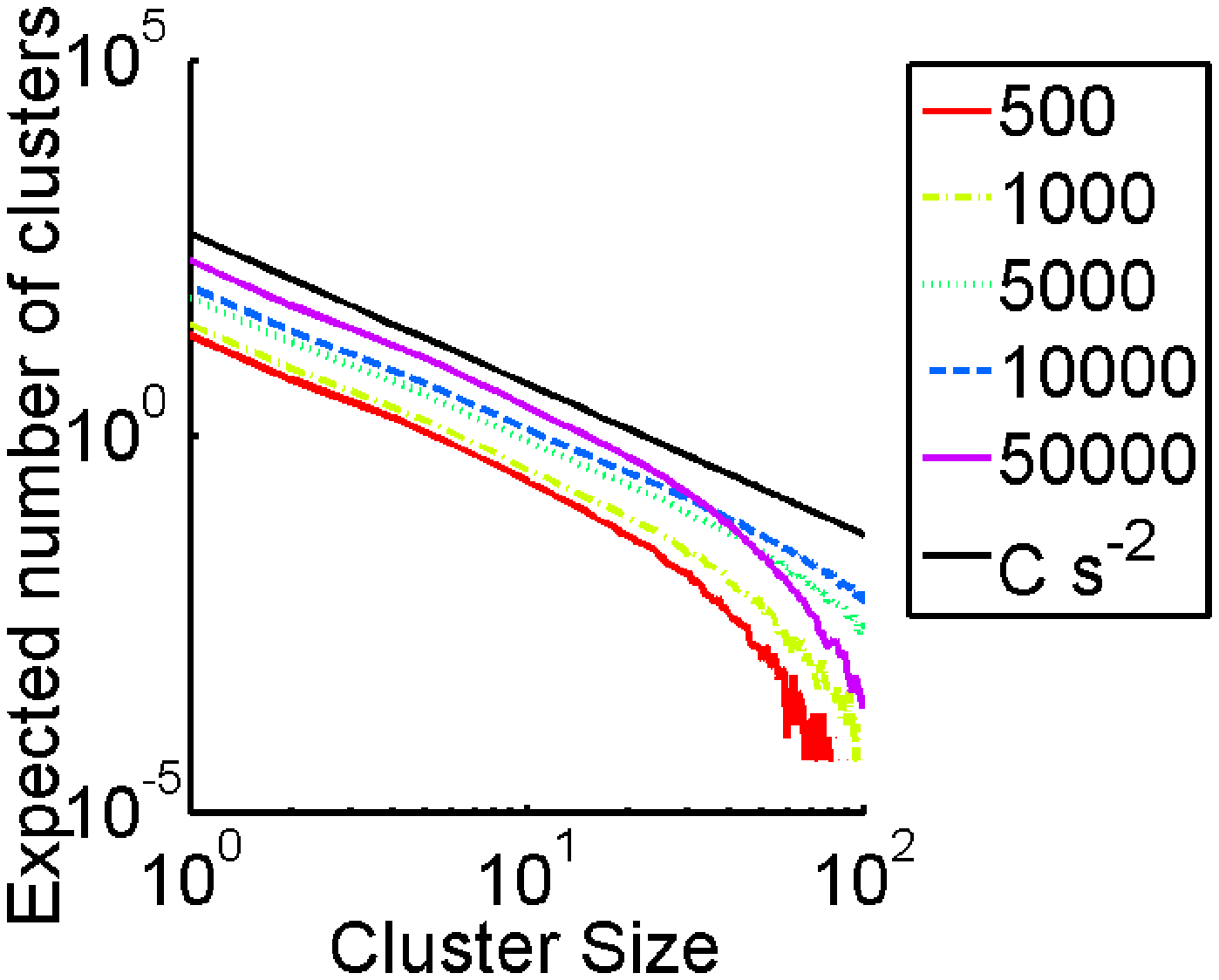}}\\
 \subfloat[]{\label{powerLawOverTime}\includegraphics[width=80mm]{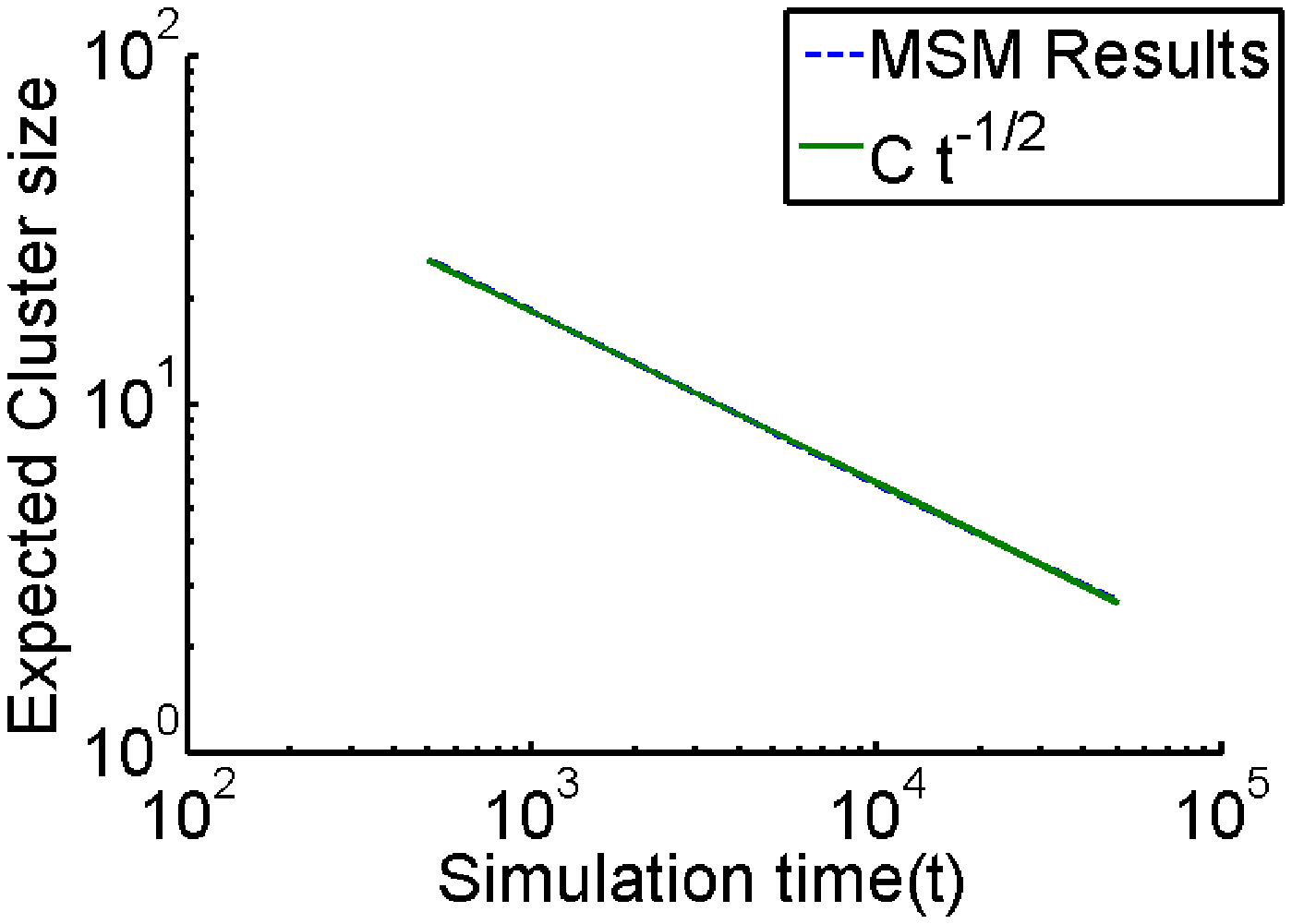}}\quad
 \subfloat[]{\label{expecClustOverDens}\includegraphics[width=80mm]{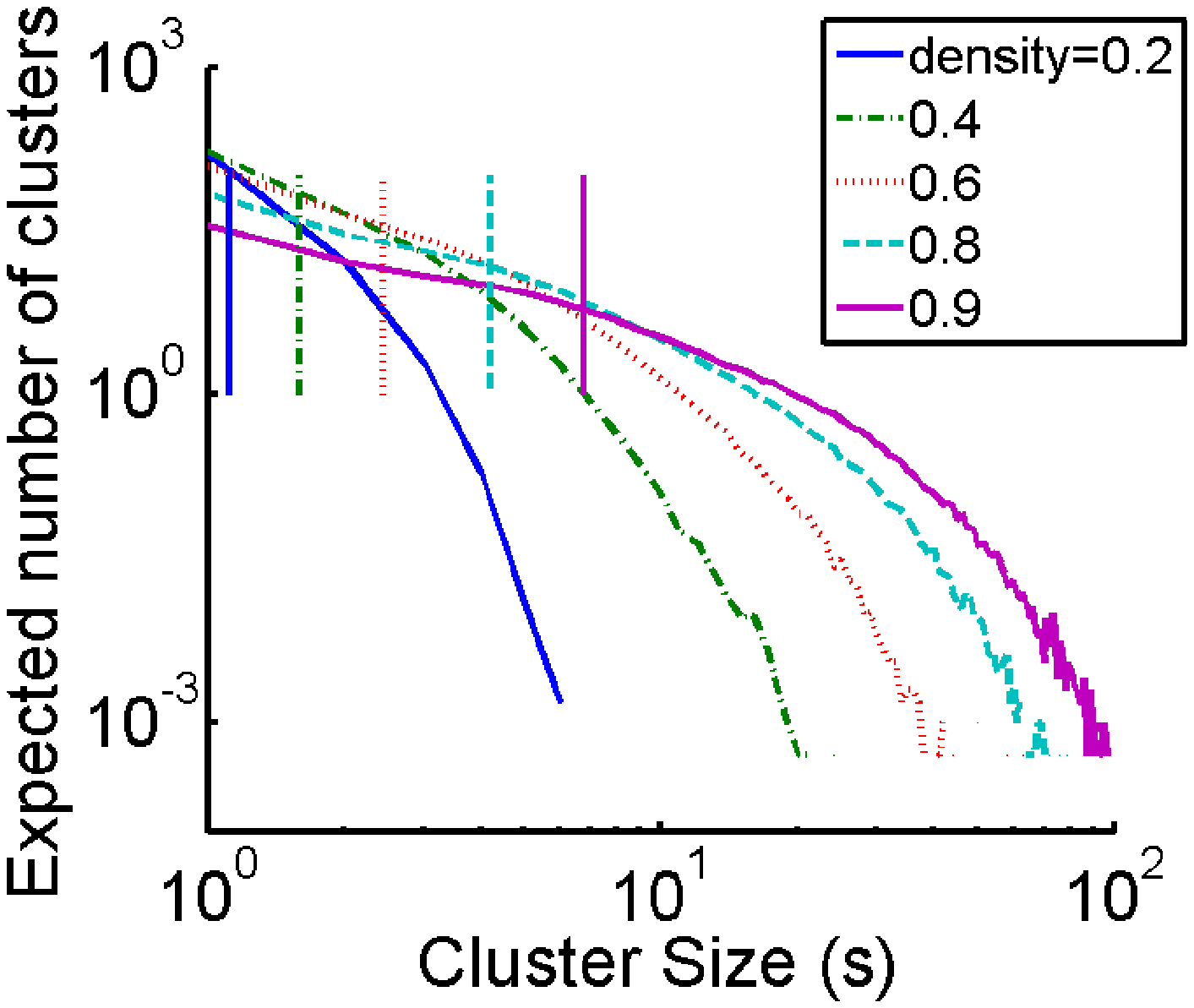}}
 \end{center}
\caption{ Results of MSM simulations of rods reversing every
T=8 performed $20,000$ times. (a)-(c) to correspond to simulations with top-hat initial conditions and (d) corresponds to simulations with initial constant average density. (a) Average cell density at
different times. (b) Expected cluster frequencies for different
times, i.e. the average number of clusters of a given size obtained
in the simulations. (c) Average cluster size over time on a log-log scale. (d) Expected cluster frequencies at end time $50,000$, after cooking initial conditions, for constant density simulations of ensemble size $2,000$, where the vertical bars signify the average cluster size.} \label{8reversalperiod}
\end{figure}

We performed multiple MSM simulations of 1D dynamics of
initially localized distributions of bacterial density and performed
ensemble averaging over these simulations. The ensemble serves to
approximate averaging over angles or the full 2D problem. We choose the
"top-hat" initial distribution (constant density around the center
of the domain and zero density to the left and to the right of the
center). Initial top-hat density profile was typically obtained by
a dense initial packing of bacteria in the domain of width 1000
around $x=0$. The typical size of a statistical ensemble was $20,000$. We
determined the cellular density (volume fraction) by calculating the
average number of times the given location was occupied (see Figure
\ref{8reversalperiod}a). Qualitatively, the cell densities spread
out symmetrically away from the center of the top hat. For later
times a steep slope develops around density $p_0=0.2$, which implies
that the rate at which the density spreads out depends on the local
cell density.
The cells' movement frequently causes them to collide with each
other. When two cells are trying to move into each other's space,
they stall (jam) until at least one reverses. This stalling, on average, shifts the
mean location of their oscillatory movement away from the location
at which they stall. If no other cells are nearby, the cells may collide
again or separate further away due to fluctuations in the mean location of their oscillatory movement. If other
cells are nearby, these outer cells have their mean location shifted
outwards while the original cells' mean locations are shifted closer
together. Through these shifts, the cells steadily spread away from
each other.

In highly packed situations, the cells cluster together. Two cells
can jam together forming two-cell cluster, and if another cell is
moving in a direction of a that two-cell cluster then it may join
the cluster forming three-cell cluster etc. To measure the amount of
clustering, we calculated the frequencies of cluster sizes at
different moments of time which is shown in log-log plots in Figure
(\ref{8reversalperiod}b) for MSM simulations with top-hat initial
conditions. It is seen that the frequency of cluster sizes follows
$-2$ power law. It is seen that the average cluster size decreases
over time as initially densely packed cells expand. Figure
(\ref{8reversalperiod}c) shows that this decrease follows a power
law decay over time with calculated exponent $-0.4965$ which is
close to $-1/2$ power laws expected from diffusion.

This suggests that during early times, many of the cells are found
in large clusters where they cannot move. At later times, the large
clusters rapidly break up into smaller clusters allowing cells to
move. Also, the individual cells near the boundaries spend a
significantly greater percentage of their time moving.

Figure  \ref{8reversalperiod}a shows that the dynamics of the
cellular density is smooth and slow compare with the velocity $v=1$
of individual cell motion. It suggests the ensemble averaged
distribution of cells  at each moment of time and at each point in
space is in statistical quasi-equilibrium. Below we study dependence
of quasi-equilibrium on local cellular density, cellular collision
times and cluster sizes.   We also performed a second type of MSM
simulations  with periodic boundary conditions and uniform average
densities to study statistical equilibrium of cellular motion.

\section{Elementary laws of collisions (jams) between cells and equilibrium motion of cells in the limit of zero noise of the reversal period}
\label{section:elementarycollisions}

If the noise in the reversal time $T$ is absent then
in a vanishing  cell density limit each cell
experiences periodic motion in space  and time with its center of
mass not moving on average (after averaging over time period $2T$).
However, the experimentally observed \cite{Welch01} small noise in the reversal time results in the random walk of the average position   (averaged over time
period $2T$) of the center of mass of each cell at time scales above $2T$. That random walk results in  collisions of cells for any finite density. As density goes to zero these collisions become
more and more rear because it takes more time for cells to span the average distance  between them through random walk. Below in this Section we consider the limit when we completely neglect that random walk, i.e. we
neglect the noise in $T$.

For nonzero density of cells, there is a finite probability for two
neighboring cells to collide (jam). By jam, we mean that one cell tries to move where another cell is located,
 but the excluded volume principle prevents them from moving. The term "jam'' in this paper is similar
to the term "collision". The subtle difference is that by
collision we mean that a cell jams with another cell with subsequent unjamming, i.e. the cell is free to move after a jam.

We distinguish two types of jams. First type is a  ``{\it pairwise
jam}". It occurs when two neighboring cells are jammed directly
because they try to move in opposite directions towards each other
but that motion is prevented by the excluded volume principle.
Second  type of jam  occurs when a cell 1 tries to move in the same
direction as a neighboring cell 2  but that cell 2 is jammed by
another cell(s) (e.g. by cell 3). We refer to such type of a jam of
cell 1 as  ``{\it indirect jam}". Such  jam  is an indirect one
because there is no direct (pairwise) jam between cells 1 and 2. A
typical example is when cell 1 moves towards neighboring cell 2
while cells 2 and 3 have a pairwise jam. After cell 1 touches cell 2
they together (cells 1,2 and 3) form three-cell cluster with
pairwise jam between cells 2 and 3 and indirect jam with cell 1. We
also say that a given cell is in a  ``{\it cluster jam}" if it is
either pairwise or indirect jam.  The pairwise collision time
$\tau_{pair}$ is always smaller or equal to $T$ because of reversal
of the direction of cellular motion. In contrast, the cluster jam
time $\tau_{cluster}$ can be arbitrary large if cells stay inside of
a large cluster.  Also in Section
\ref{section:Analyticalapproximation} we use total jam time $\tau$
per period $T$, i.e. the time during which a given cell remains
jammed (either directly or indirectly) per period $T$. With such
definition $\tau$ never exceeds $T$.

 Assume that density is small so that mostly pairwise jams occur. In
 such a case jamming of two cells lasts until one of the cells
 reverses. After that they move together in the same direction until
 the second cell reverses. After the second cell reversal, cells move
 in opposite directions away from each other. Assume that all other
 cells are still far away. Then after the first cell reverses for a
 second time both cells will move in the same direction, and after
 the second reversal of the second cell they will move towards each
other. Exact calculation  shows that these two cells will never jam
again in the absence of other cells. Instead, exactly at the moment
when these two cells touch each other, the first cell will reverse
for a third time and they will move in the same direction again.
This pattern of periodic motion without jamming of these two cells
will continue for arbitrary long time (or until another, third cell,
would approach them close enough to jam with one of these two
cells). It means that any two isolated cells jam only once and after
that both cells will experience periodic motion without disturbing
one another.

A similar interaction pattern occurs if we consider a system of
three or more cells moving in an infinite spatial domain. After
several collisions (jams) between these three or more cells, they
will also end up in the state where they will not jam any more and
all cells will experience periodic motion without touching each
other. A center of mass of each cell participating in a jam shifts
relative to its average position at a distance $v\tau_{pair}$ to the
left or to the right (depending on which side it has a jam), where $\tau_{pair}$ is the collision time. However, after all collisions
are over, center of mass of each cell experiences periodic motion
and no average motion (averaged over the period $2T$) is observed.
We refer to such state as an equilibrium motion of Myxobacteria.
Note that equilibrium motion is quite different from the equilibrium
distribution (Gibbs distribution) in statistical mechanics
\cite{LandauLifshitzStatisticalPhysics1980} because Myxobacteria are
always self-propelled and are not subject to any type of thermal
equilibrium. Starting with a finite number of initially densely
packed Myxobacteria, after finite number of collisions and provided
Myxobacteria divisions are neglected, the bacterial colony expands
to such size that there will be no more collisions between cells.
After that the average size of the colony remains the same with
bacteria moving periodically at equilibrium.

We now calculate the density of Myxobacteria $p_0$ at which a
transition occurs from motion with collisions to equilibrium motion.
First, consider two neighboring cells and assume that they have
phases $\phi_1$ and $\phi_2$, respectively. Generally $-2T \le
\phi_1-\phi_2\le 2T$ but assuming periodicity over time $2T$ we can
always add a multiple of $2T$ to each of the phases, $\tilde
\phi_j\equiv\phi_j+n_j2T$, $j=1,2$ ($n_j$ are integers), to keep the
difference of modified phases inside a twice smaller interval:  $-T
\le \tilde \phi_1-\tilde \phi_2\le T$. For $p=p_0$ cells do not jam
but during a part of the time interval $2T$  they move together
(attaching to each other) in the same direction until one of them
reverses. After that, they move in opposite directions from each
other for the time interval $|\tilde \phi_1-\tilde \phi_2|$. After
that second cell reverses and both cells move in the same direction
etc. Minimum separation between centers of mass of these two cells
is $L$ and maximum separation is $L+2v |\tilde \phi_1-\tilde
\phi_2|$. So the distance $L_{dist}$ between the average positions
of centers of mass is $L_{dist}=L+v |\tilde \phi_1-\tilde \phi_2|$.
Now, to calculate the average density $p_0$ of many cells we average $L_{dist}$ over phase differences $0\le |\tilde \phi_1-\tilde
\phi_2|\le T$ resulting in the critical density
\begin{eqnarray}\label{p0def}
p_0\equiv \frac{L}{\langle L_{dist}\rangle}=\left [T^{-1}\int^T_0 (L+v\phi)d \phi \right ]^{-1}=\frac{L}{L+vT/2}.
\end{eqnarray}
For the standard values $v=L=1, \ T=8$ it yields that $p_0=0.2.$

If initially there is a localized distribution of cells with
the average density $p>p_0$, then these cells will spread out with
collisions until their density reaches $p=p_0.$ If initially
$p<p_0$, then some  redistribution of cellular density may occur
when the average distance between centers of mass of two neighboring
cell is $L_{dist}<L+v |\tilde \phi_1-\tilde \phi_2|$. Because
average density is low, this would result only in a local
redistribution of the positions of cells without much change in the
macroscopic cellular density. After initial spreading out no collisions or cellular density transport will be observed.

We  conclude that in
order to observe transport of a system of self-propelled rods without noise in the reversal period $T$ at
long times one needs to incorporate in the model a source of the
density gradient. In Myxobacteria swarms, such a source is present
because of division of cells in the center of the bacterial colony.
Thus any transport of self-propelled rods without  noise in  $T$  is a collective phenomena  with the threshold density $p_0$ required for transport.

\section{Re-introduction of the noise of the reversal period in the small density limit}
\label{section:reintroductionofnoise}

We now take into account the noise of the reversal period $T$ in the analysis of the previous Section.
In that case collisions between cells occurs even for $p<p_0$ because random walk of the average position of cells
causes them to move at arbitrary large distance until finally colliding with other cells. When density approaches to zero the frequency of collisions also goes to zero.
But if $p\to p_0$ from below then cells collide typically at each period $2T$ with the collision time $\sim \Delta T_0$ (so that for $\Delta T_0\to 0$ that collision time would vanish).
Thus $p_0$ separates two regime of collisions: for $p<p_0$ rare collisions primary occur because of the noise of $T$ while for $p>p_0$ collisions are dominated by frequent collisions.
At the transition densities $p\sim p_0$ the contributions of both of these effects are comparable.

Thus a transport of Myxobacteria is a mixture of two effects. First effect is  the diffusion of individual cells due to the noise in the reversal period $T$ which dominates for small densities $p<p_0$.
Second effect is due to the frequent collision of cells during each period $2T$ making that effect essentially collective one.
These two regimes make Myxobacteria quite distinct from bacteria like \emph{E. Coli} or amoeba
\emph{Dictyostelium discoideum} which experience diffusion as random walk of Brownian-like particles \cite{Alber06,Alber07,Lushnikov08} without any periodic motion.

\section{Multiple collisions and cell clustering for large cellular densities}
\label{section:multiplecollisions}

\begin{figure}
\begin{center}
\subfloat[]{\label{jamExamplev1}\includegraphics [width=80mm]{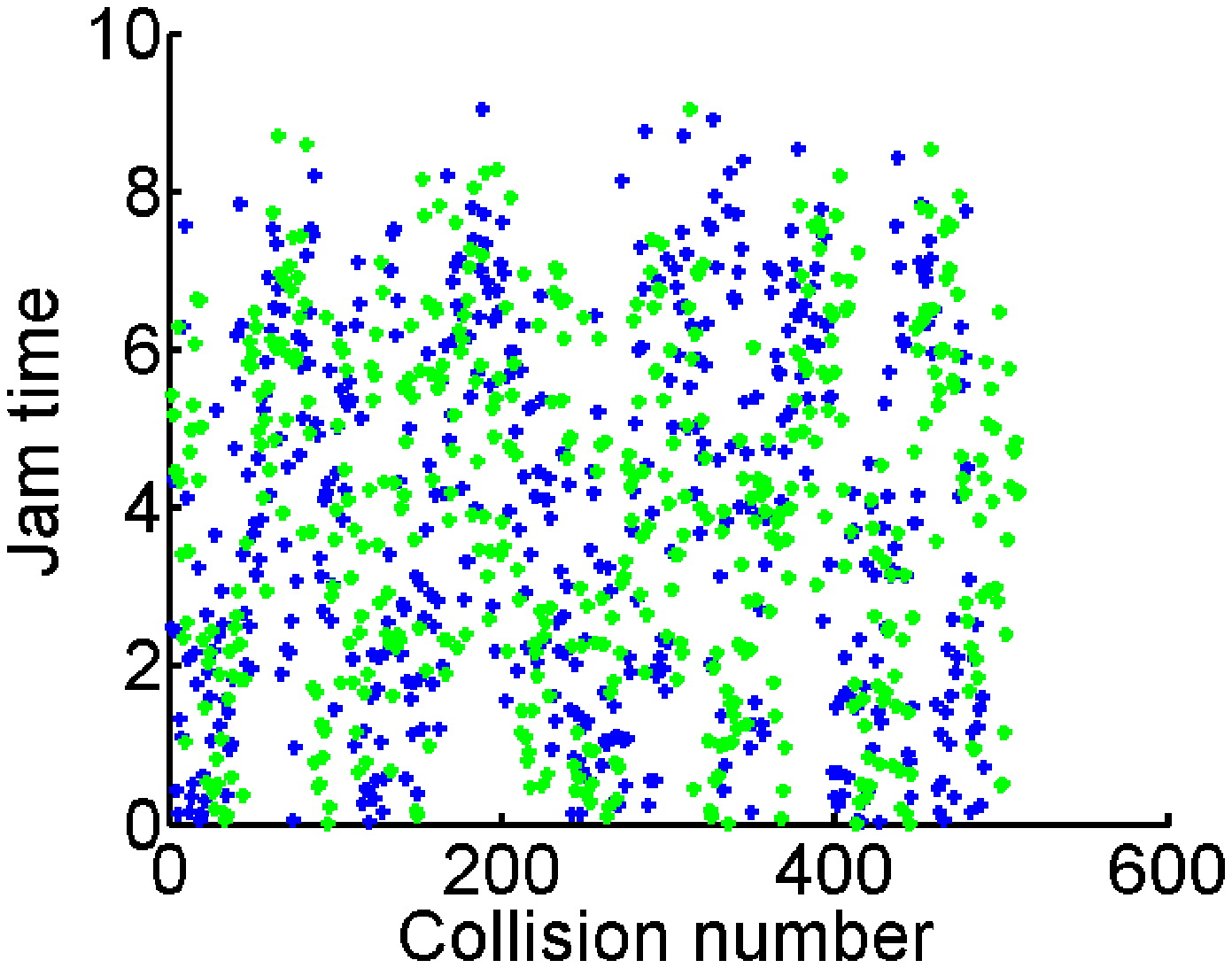}}\quad
\subfloat[]{\label{jamExamplev2}\includegraphics [width=80mm]{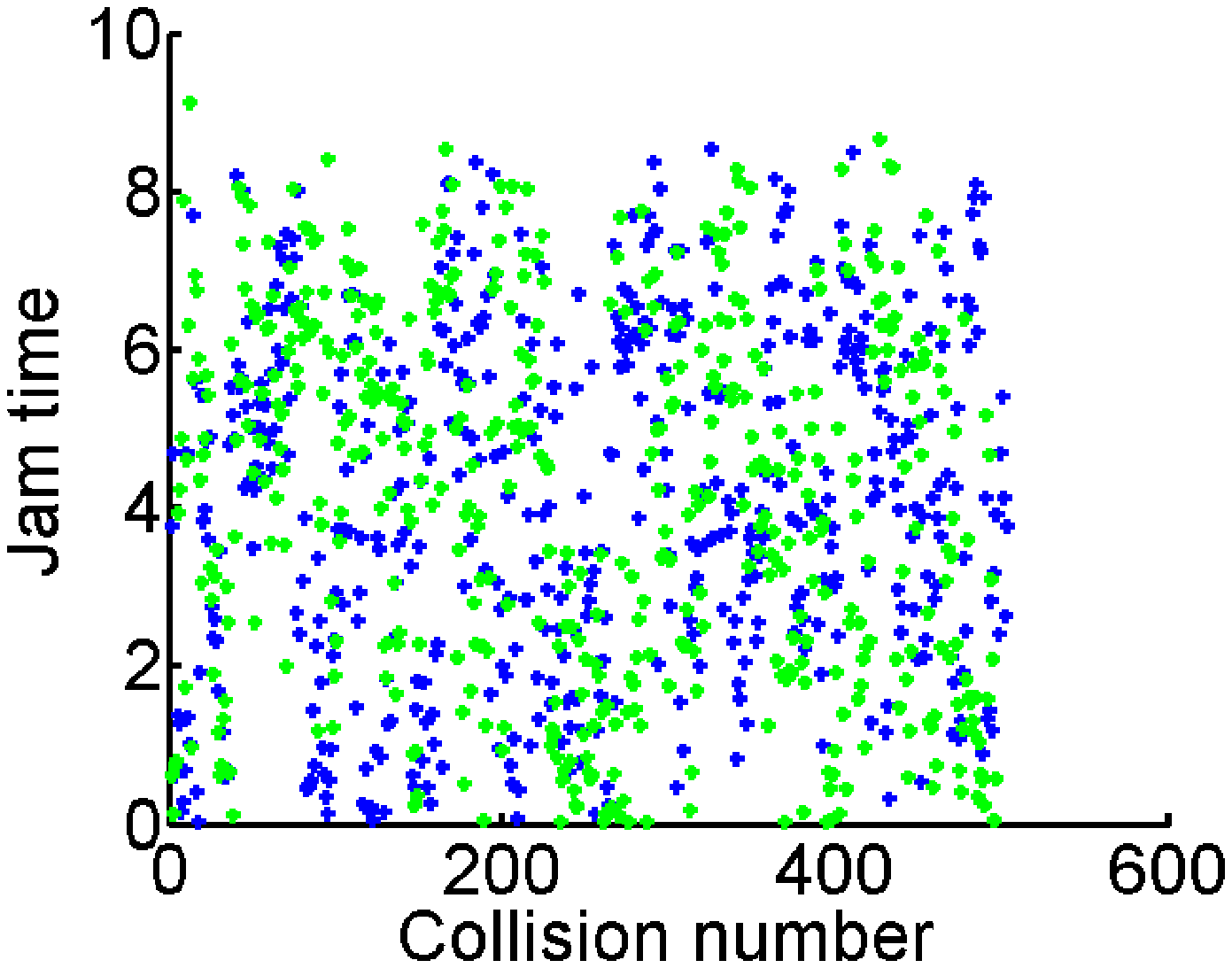}}\quad\\
\subfloat[]{\label{jamExamplev3}\includegraphics [width=80mm]{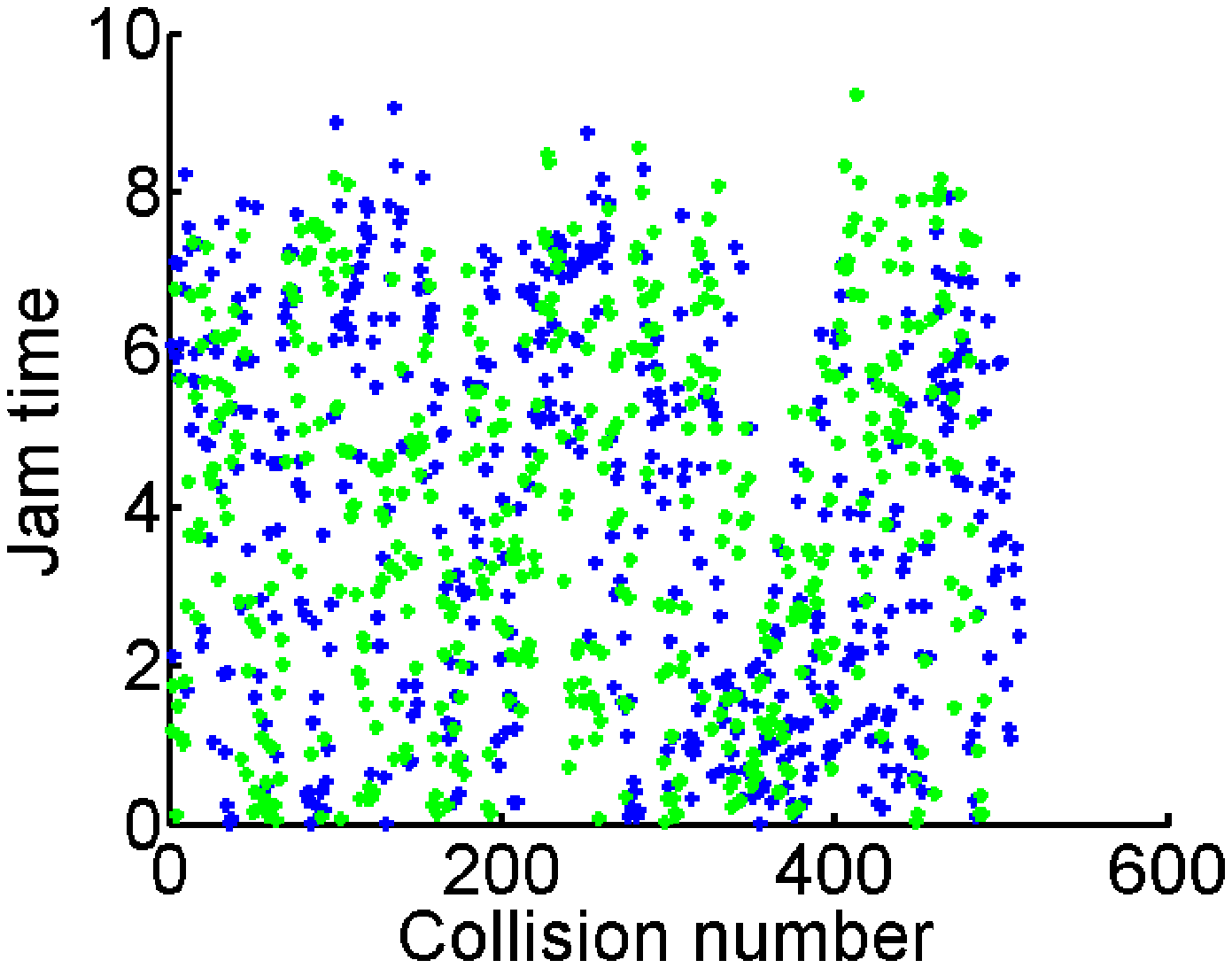}}\quad
\subfloat[]{\label{jamExamplev4}\includegraphics [width=80mm]{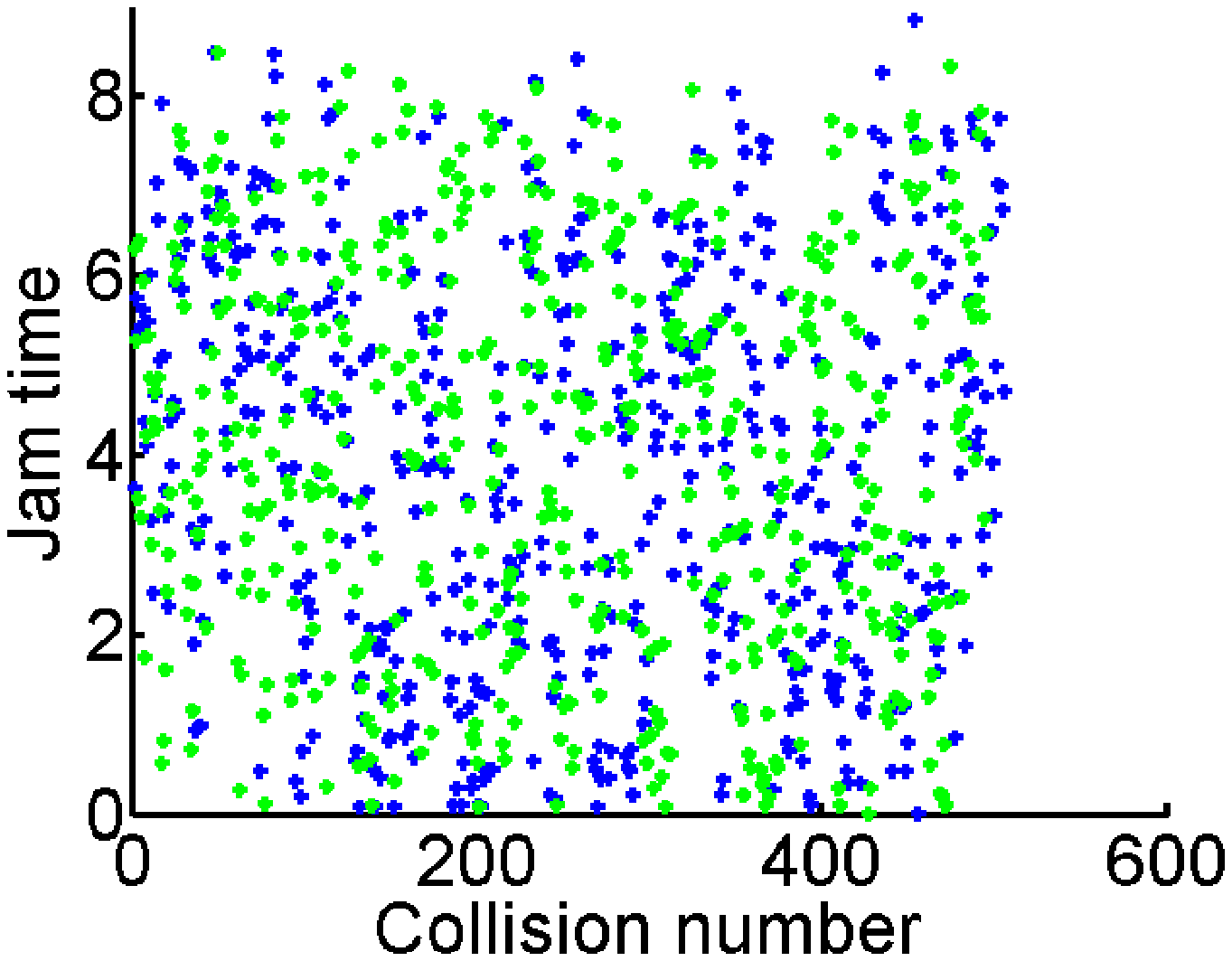}}\quad
\end{center}
\caption{(a-d) Four stochastic realizations of the pairwise jam time that a single cell experiences with its two adjacent neighbors. Collisions with the left and right neighbors are colored blue and green respectively for a total of 1,000 different collisions at average cellular density 0.95 and $\triangle x=0.001$.} \label{jam}
\end{figure}

If the cellular density $p$ is not small ($p>p_0$) so that cells
typically experience collisions during each period $2T$, then cell
motion is more complicated than described in Section
\ref{section:elementarycollisions} which is based on rare pairwise
collisions. In addition we assume here and below that there is nonzero noise in $T$ as in Section \ref{section:reintroductionofnoise}.
Figures \ref{jam}(a-d) show  pairwise jam time (jam
duration) versus collision number that occur between three adjacent cells for the
average cellular density $p=0.95$. In that case cells occupy 95\% of
total volume and between two reversals each cell can cover up to
$vT/L=8$ cell volumes meaning that it could collide with multiple
cells if allowed to.
It is shown in Figure \ref{jam} that distribution of pairwise
jam time $\tau_{pair}$ is random. Such regime typically occurs
closer to the bacterial colony center where cell flux caused by cell
divisions is large and it keeps the system far from equilibrium motion
state as described in Section \ref{section:elementarycollisions}.
There are at least two situations where such a far from equilibrium
state is possible. First is the high density gradient case caused by
bacterial division (as mentioned above). The second case occurs if
no-flux boundary conditions maintain a large density of Myxobacteria
in a domain with fixed volume. In both cases the rate of bacterial
jamming is high and the collision times are randomly distributed.

\begin{figure}
\begin{center}
\includegraphics [width=90mm]{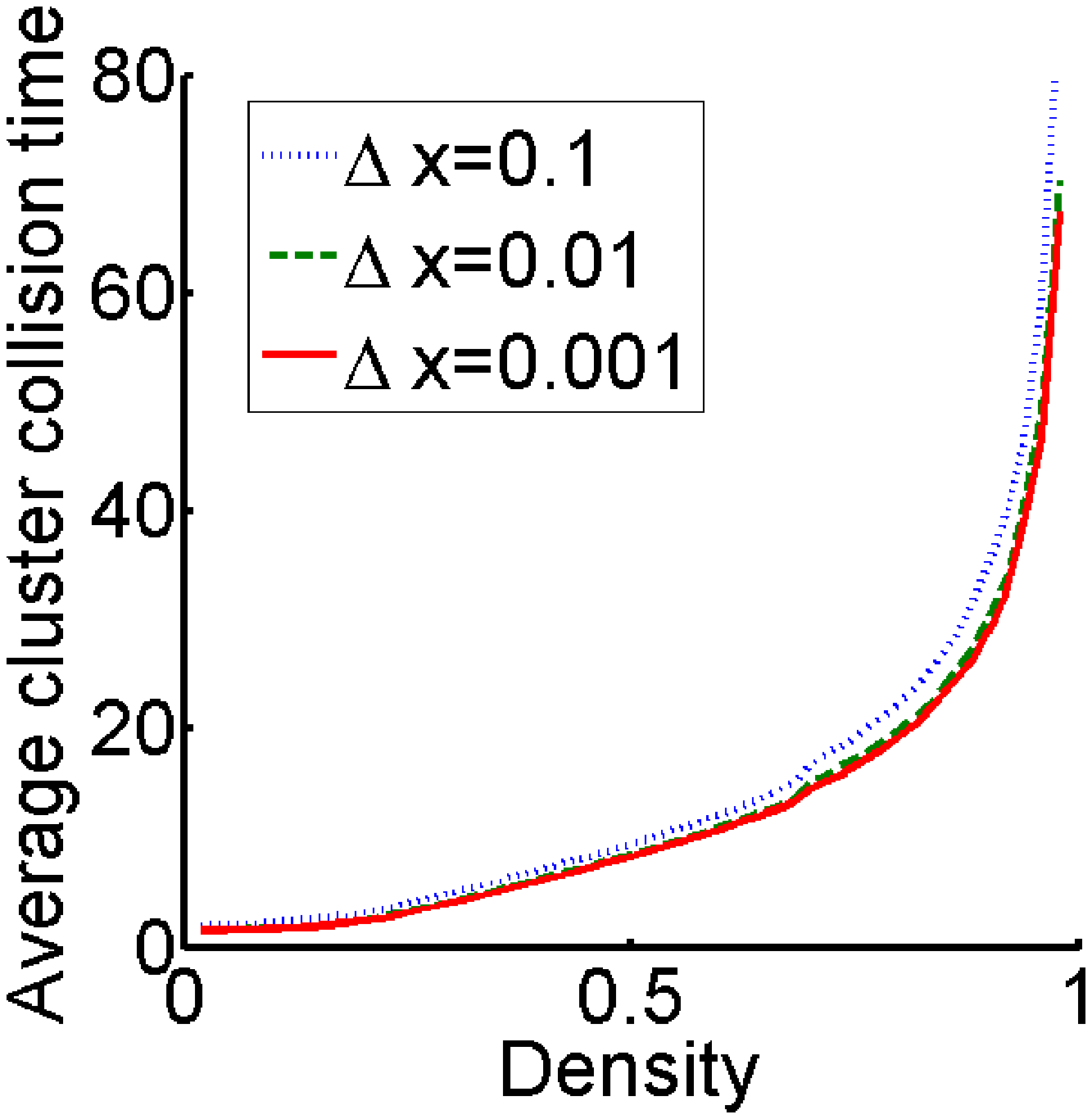}
\caption{Average cluster collision time $\tau_{cluster}$ as a
function of density $p$ for different values of $\Delta x$. It is seen that  $\tau_{cluster}\to \infty$ for $p\to 1.$}
\label{fig:clusterjam}
\end{center}
\end{figure}

Another effect which occurs in case of large densities is the high
probability of formation of clusters consisting of more than two
bacteria. As density of bacteria approaches one, all bacteria jam in
large clusters. Unjamming bacteria from large cluster might take a
lot of time because leftmost or rightmost bacteria in the cluster
need to move away providing space for the bacteria in the center of
the cluster to move in. As a result, many cells stay jammed in a
cluster for a long time for large densities. Figure
\ref{fig:clusterjam} shows that the average cluster collision time
$\tau_{cluster}$ (averaged over ensemble of MSM simulations)
diverges for $p\to 1.$ Values of $\tau_{cluster}$ for different
$\Delta x$ show good convergence to the continuous limit $\Delta x\to 0$.
E.g., curves for $\Delta x=0.01$ and $\Delta x=0.001$ are almost indistinguishable.

Figure \ref{8reversalperiod}b shows the distribution of
cluster sizes for MSM simulations starting from a top-hat initial
condition. Figure \ref{8reversalperiod}d shows the cluster size
distribution for different densities $p$ for MSM simulations with
uniform average density and periodic boundary conditions (second
type of MSM simulations is described in Section
\ref{section:MSMSimulations}.

\section{Macroscopic Nonlinear Diffusion Model and MSM simulations}
\label{section:MacroscopicNonlinearDiffusionModelvsMSMsimulations}

\subsection{Nonlinear Diffusion Model and Its Limitations}
\label{section:MacroscopicNonlinearDiffusionModelvsMSMsimulationsSubsectiona}

If collisions are frequent and, additionally, the distribution of
the reversal phases and initial position of cells are random, then
we assume that collective dynamics of cells is diffusion-like and it
is described by the equation of the general type
(\ref{nonlineardiffusion1}). In this section, we  obtain a numerical
approximation of the diffusion coefficient $D(p)$ in
(\ref{nonlineardiffusion1}) to match the results of MSM simulations.
This is achieved by running MSM simulations with a top-hat initial
distribution and using Boltzmann-Matano (BM) analysis
\cite{BoltzMat} applied to the ensemble-averaged MSM density profile
on the right half of the spatial domain at time $t=t_D$. Here and
below $t_D$ describes the time at which we apply BM analysis.
(Description of the BM analysis is given in Appendix A). Then we
demonstrate that numerical solutions of (\ref{nonlineardiffusion1})
with $D(p)$ obtained using BM analysis, yield density dependence on
space and time which are in a very good agreement with the one
obtained using MSM simulations, justifying initial assumption
of collective dynamics of cells being diffusion-like.

Results of MSM simulations unavoidably have noise due to the finite
size of stochastic ensemble used to determine cellular densities. BM
analysis relies on calculating derivatives of density and we apply
Gaussian filter to the MSM density data to smooth out both $p$ and
all its derivatives \cite{nixon2008feature}. We checked that the change of parameters of the
Gaussian filter resulted in only small corrections without any
systematic error.

Typically, to perform BM analysis we run MSM  using an initial
top-hat distribution of length $1,000$ in a domain of size $4,000$
(see the end of Section
\ref{section:MicroscopicStochasticModelcellmotion} for more
details). For most of our simulations, the top-hat is wide enough so
that during simulation time the density at the middle of the domain
remains close to the initial density $p_{max}$ (in most simulations
$p_{max}=1$). It means that cells mostly move near the boundary of
initial top-hat distribution while at the middle of the domain the
cellular density is almost constant. This allows us to ignore the
left half of the domain and treat the cell distribution as if it
were step-wise shaped in an infinite domain. This is necessary in
order to perform BM analysis which is exact for infinite spatial
interval with step-wise initial conditions only. In Appendix B we
study the accuracy of BM analysis for a finite width of top-hat
initial conditions.

Another limitation of the BM analysis is that it requires
calculating $(dp(x)/dx)^{-1}$. Due to the presence of the regions
where the density is constant, singularities of  $(dp(x)/dx)^{-1}$
may be generated in the calculation of the non-linear diffusion
coefficient near the end of the diffusion curves where $p$ is close
to 0 or $p_{max}$. These artificial singularities result from a loss
of numerical precision near singularity of $(dp(x)/dx)^{-1}$ which
is clearly seen near $p=0$ and $p=1$ in all Figures below that
include $D(p)$. To reduce such loss of numerical precision we only
perform BM analysis in the neighborhood of the interface that
encompasses the initial step of a top-hat instead of the entire
right half of the domain. It can be also mitigated by performing a
cubic spline interpolation of $D(p)$ from the domain $0<p<1$ to
values around $p=0$ and $p=1$. This, however, appears to be not
necessary because the loss of precision does not affect prediction
of the density dynamics in (\ref{nonlineardiffusion1}) in any
significant way.

\begin{figure}
\begin{center}
 \subfloat[]{
 \includegraphics[width=80mm]{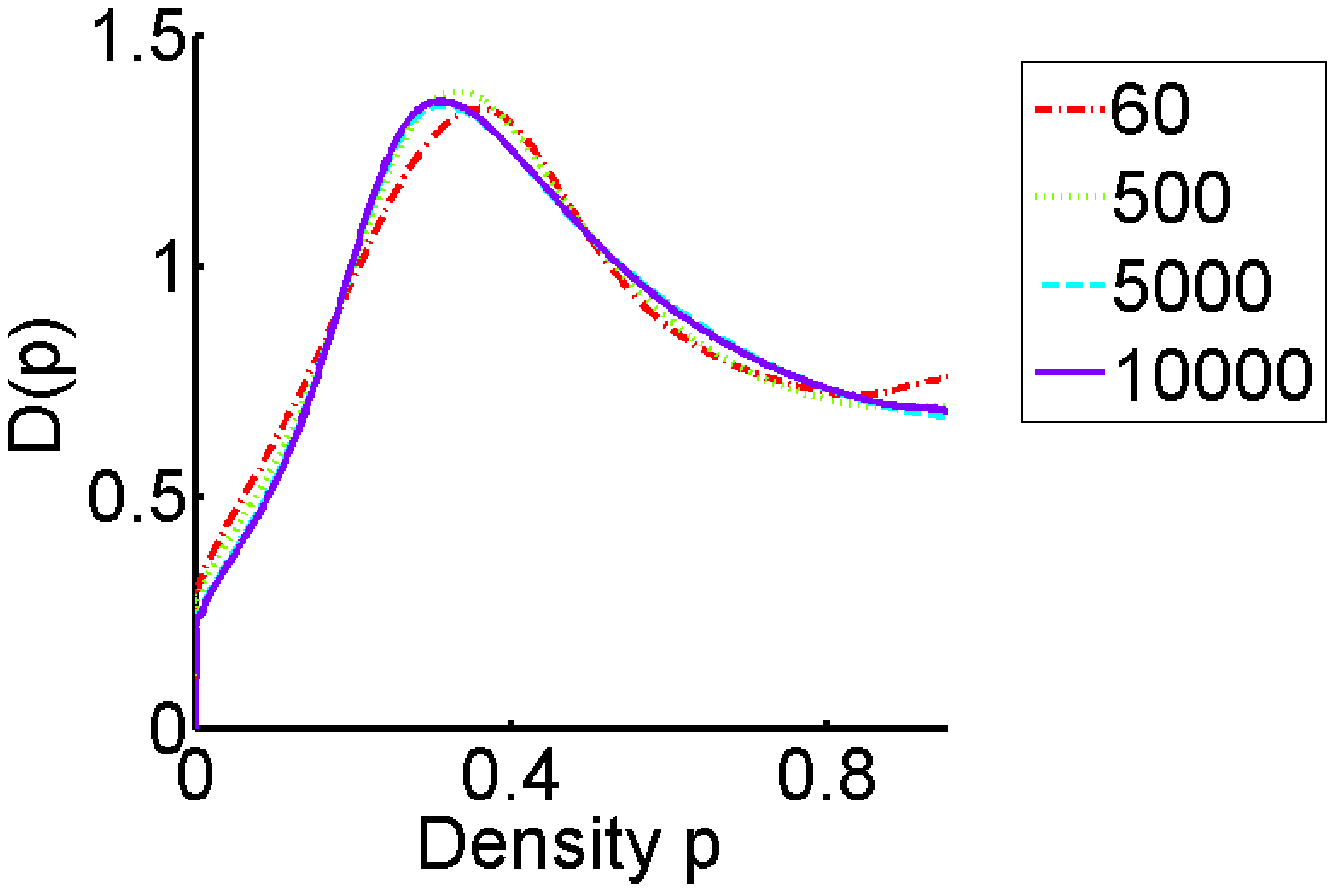}
}\quad
 \subfloat[]{
 \includegraphics[width=80mm]{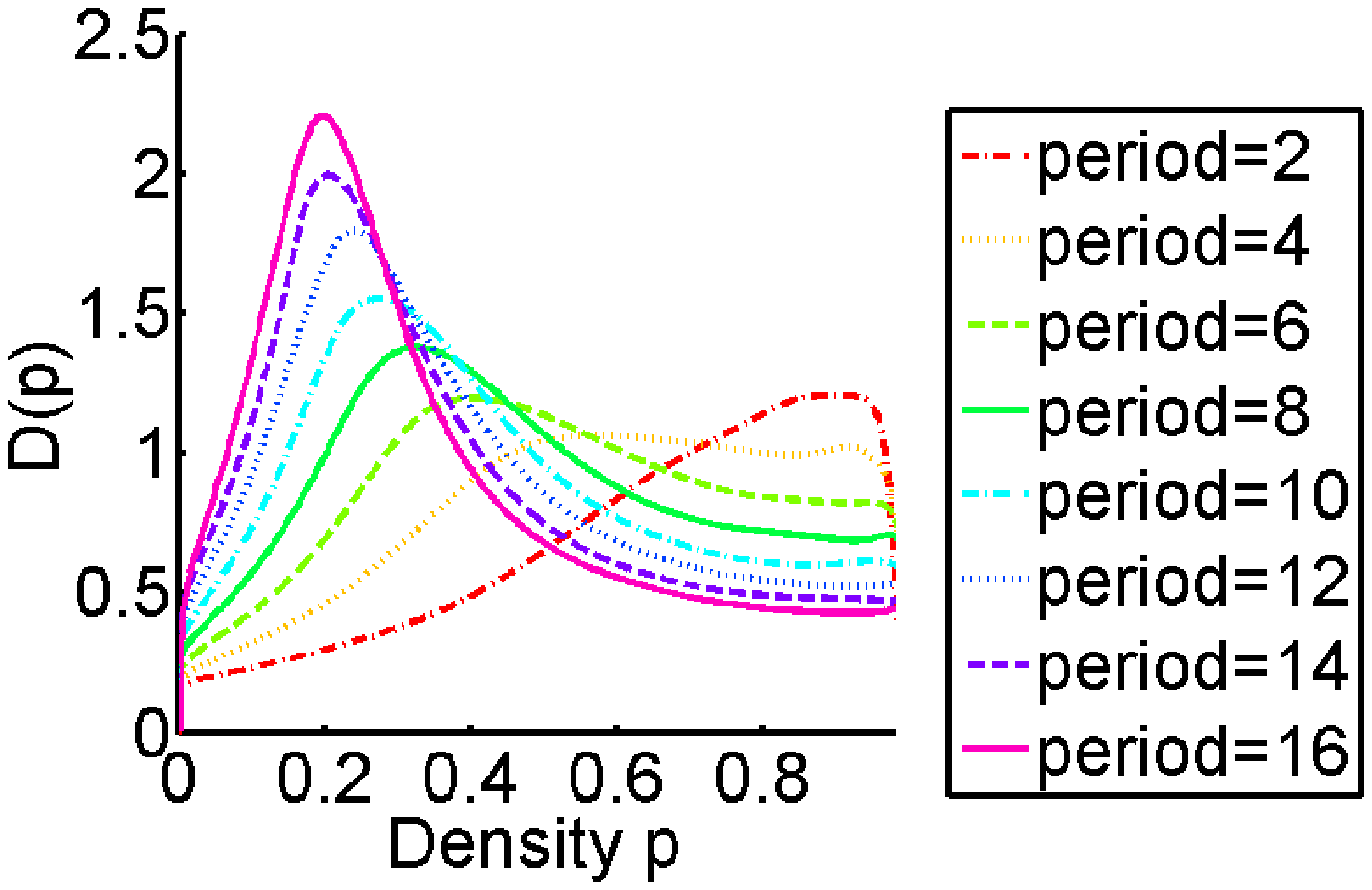}
}\\
 \subfloat[]{
 \includegraphics[height=50mm]{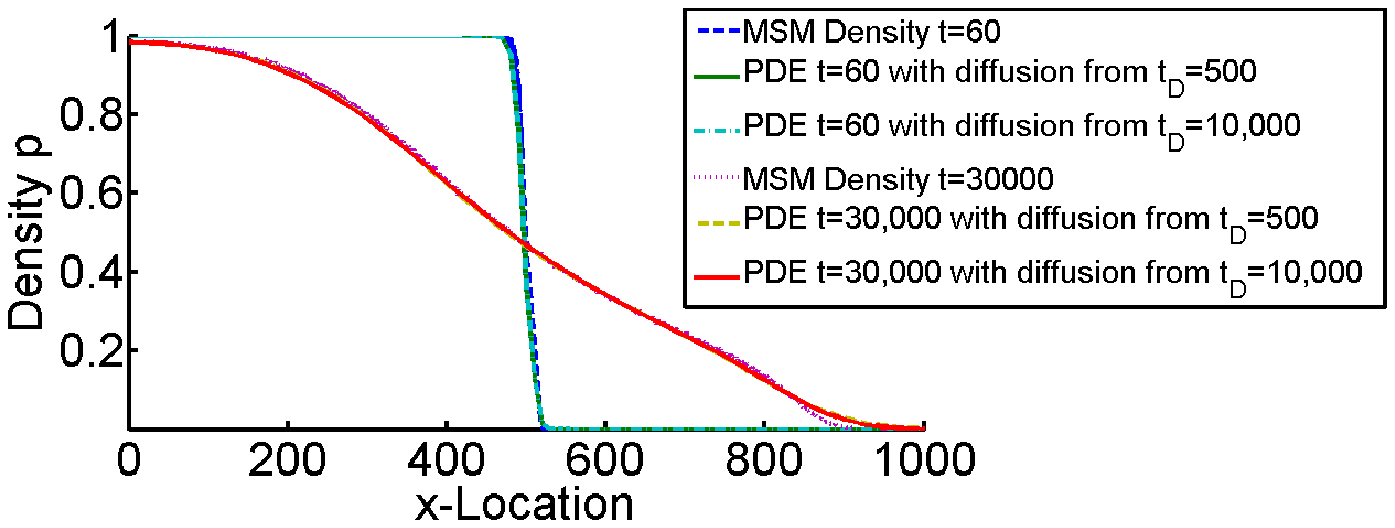}
}
\end{center}
\caption{(a) Diffusion coefficient $D(p)$  generated by BM analysis at different times
$t_D$  for $T=8$.  (b) $D(p)$  generated by BM analysis for varying
reversal periods $T$ for diffusion curves calculated from BM analysis at $t_D=500$. (c) Density profiles from MSM simulations vs. PDE density
profiles (solutions of  (\ref{nonlineardiffusion1})) at $t=60$  and $t=30,000$ obtained using diffusion coefficients from (a)
with $t_D=500$ and $t_D=10,000$. It is seen that the fit between two types of density profiles is very good except the region of small density $p<p_0=0.2$.
 } \label{diffusionCurves}
\end{figure}
Since the BM analysis approach for calculating the diffusion
coefficient assumes that the nonlinear diffusion equation
(\ref{nonlineardiffusion1}) is solved on an infinite domain, there
will be errors in calculating the diffusion coefficient if there is
significant density near the boundaries of our finite computations
domain. So if were to choose $t_D$  that is too large, the analysis
would fail. Also, if were to choose $t_D$ that is too small, then
not enough cells would reverse to generate diffusion. We found that
any time between $t_D=125$ and $t_D=10,000$ appears to be sufficient
for generating reasonably universal diffusion curves for $T=8$ as
shown in Figure \ref{diffusionCurves}a. In other words, diffusion
coefficient curves $D(p)$ generated at different times $t_D$, are
close to each other. This near-independence of  $D(p)$ from $t_D$
justifies our assumption of the collective dynamics of cells being
diffusion-like and described by (\ref{nonlineardiffusion1}).

Small difference between different diffusion curves $D(p)$ generated
at different times $t_D$, is due to the absence of diffusion for
$p<p_0$, where the critical density $p_0$ is defined in
(\ref{p0def}). For $p<p_0$ there are  practically no jams between
cells (see Section \ref{section:elementarycollisions}) and,
subsequently, there is no diffusion. (Some occasional jams are still
possible only because cells are not in fully quasi-equilibrium and
because of finite value of $\triangle x$.)  BM analysis assumes that
$D(p)$ is an analytical function. But a sharp drop of diffusion to
zero for $p<p_0$ breaks that analyticity.  Therefore, in BM analysis
approach a drop of $D(p)$ from finite value at $p\to p_0+0$ to the
zero value at $p\to p_0 -0$ is replaced by a smooth function shown
in Figures \ref{diffusionCurves}a and b  (see also Section
\ref{section:Analyticalapproximation} for more relevant discussion
on this subject). Thus, nonzero values of $D(p)$ for $p<p_0$ in
Figures \ref{diffusionCurves}a and b are artifacts of the BM method.
Also, finite values of  $\triangle x$ contribute to nonzero value of
$D(p)$  for $p<p_0$ but that contribution is small for $\triangle x
\le 0.1$ in comparison with contribution of BM analysis. In the
limit $\triangle x \to 0$  there is no average macroscopic motion of
cells for $p<p_0$. MSM simulations indicate this by the sharp drop
of density for $p<p_0$. (Such drop is not completely vertical
because of finite sizes of cells in the MSM.) In addition,
macroscopic averaging requires taking into account a finite spatial
width of that sharp drop of density.   We conclude that the
difference between MSM and PDE curves in Figure
\ref{diffusionCurves}c for $p<p_0$ is due to the limitation of
applicability of BM analysis for $p<p_0.$

Unless otherwise specified
below we use $t_D=500$ to generate the diffusion curves.

 To demonstrate that there is little difference in the solutions of
(\ref{nonlineardiffusion1}) with diffusion coefficients $D(p)$
chosen based on BM analysis with $t_D=500$ versus $t_D=10,000$, we
compare the resulting numerical solutions with the densities
obtained using microscopic stochastic model simulations (see Figure
\ref{diffusionCurves}c). Figure \ref{diffusionCurves}c shows that
PDE density profiles $p(x)$ are almost indistinguishable for both
values of $t_D$. Furthermore, the difference between the numerical
solutions of the nonlinear diffusion equation and stochastic
simulation results are negligible except for the region $p<p_0$.

Diffusion curves for different reversal periods $T$ were also
calculated (see Figure \ref{diffusionCurves}b). Large reversal
periods $T$  produce high diffusion at low densities, and low
diffusion at high densities. Small reversal periods produce low
diffusion at low densities and high diffusion at high densities. In
the former case the cells move left or right until they collide and
they stay jammed for a long time. In the later case, the cells
rapidly oscillate left and right. Once the cells spread out,
collisions become infrequent.

\begin{figure}
\includegraphics[width=160mm]{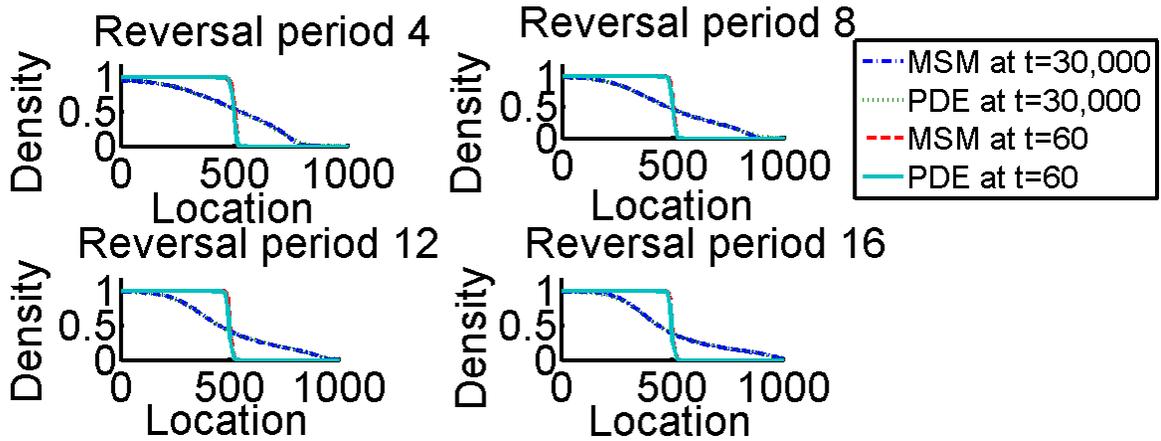}
\caption{Numerical solution of the diffusion equation (\ref{nonlineardiffusion1}) (with $D(p)$ from BM analysis) plotted against the ensemble-averaged MSM
results at $t=60$ and $t=5,000$ for different reversal periods $T$. PDE and MSM results are almost indistinguishable.}
\label{PDEDensityReversals}
\end{figure}

\subsection{Testing accuracy of the nonlinear diffusion model and macroscopic limit of MSM}
\label{section:MacroscopicNonlinearDiffusionModelvsMSMsimulationsSubsectionb}

To test that the diffusion curves in Figure \ref{diffusionCurves}b
actually predict the diffusion in the stochastic model, we compared
numerical solutions of the diffusion equation
(\ref{nonlineardiffusion1}) with $D(p)$ derived using BM analysis of
MSM simulations with different reversal periods at an early and a
later times (see Figure \ref{PDEDensityReversals}). Very good match
is demonstrated for $p>p_0$, where the critical density $p_0$ is
defined in (\ref{p0def}).

To test whether the dynamics of the discrete stochastic system is
consistently well approximated by the diffusion equation,
independently of initial density, we compared the numerical
solutions of the diffusion equation with the MSM simulations for
$T=8$ reversal period and different initial conditions (different
amplitudes of densities in the initial top-hat profile).
\begin{figure}
\begin{center}
\includegraphics[width=120mm]{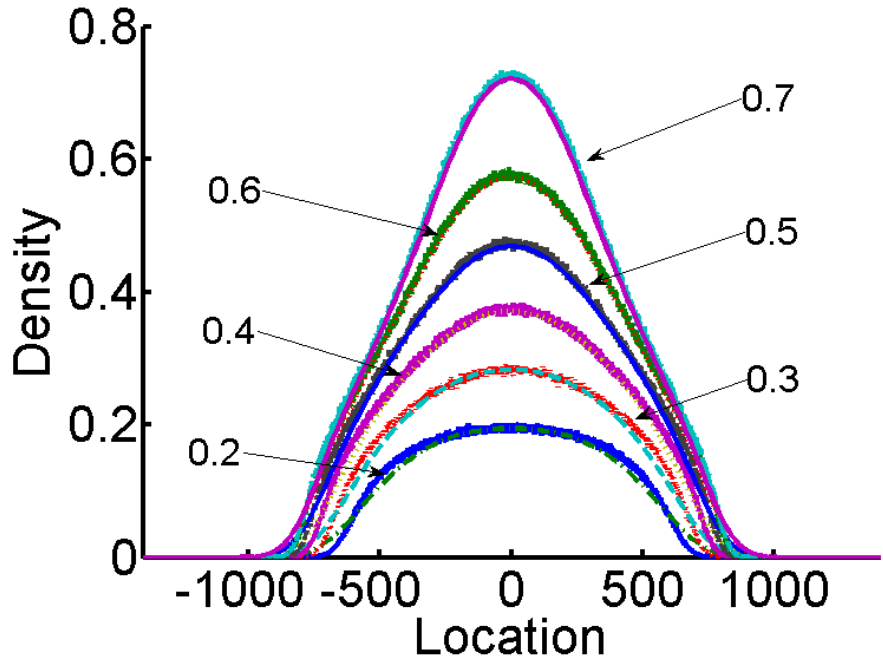}
\end{center}
\caption{Comparison of ensemble-averaged MSM simulations with numerical solution of (\ref{nonlineardiffusion1}) (with $D(p)$ from BM analysis)  using different initial densities.
MSM densities are not smooth because of finite size of statistical ensemble. It is seen that PDE results are  well approximations for MSM for $p>p_0$ if we average over fast fluctuations of MSM densities. Simulations results were taken at $t=50,000$}\label{differentPDEInit}
\end{figure}
We first generated random initial conditions with constant average
density and periodic boundary conditions and allowed cells to move
in the MSM simulations  for $t=50,000$ to reach statistical
equilibrium. After that we inserted the obtained equilibrium
distribution as a top part of top-hat initial condition and run MSM
simulations starting with these spatially nonuniform initial
conditions. Figure \ref{differentPDEInit} shows a very good match
between these simulations and numerical solutions of the nonlinear
diffusion equation. Matching is not as good for smaller densities
due to the qualitative change of diffusion and lack of collisions for $p<p_0$ as was explained above.

\begin{figure}[thp]
\begin{center}
\includegraphics[width=80mm]{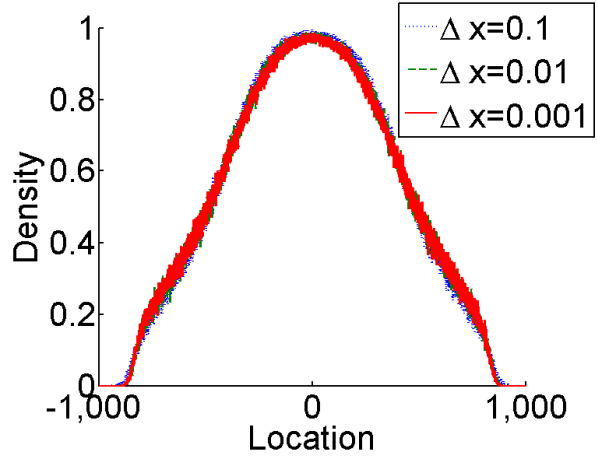}
\end{center}
\caption{Spatial distribution of density at $t=30,000$ from the ensemble-averaged MSM simulations with different $\Delta x.$ Initial condition had a form of top-hat distribution of width 1000.
} \label{densitydeltax}
\end{figure}

Since cells move on a discrete grid at discrete time steps, we test
convergence of the system to a continuous description of cell
movement by decreasing the grid spacing $\triangle x$, and by
scaling the lengths and time steps appropriately. The results of the ensemble-averaged MSM simulations  for
the spatial profile of density are shown in Figure \ref{densitydeltax}.  It is seen that the  reduction of $\triangle x$
from 0.1 to 0.001 results only in small changes in the cellular
density dynamics with density
curves for different $\triangle x$  practically indistinguishable
in   Figure \ref{densitydeltax}.  It suggests  $\Delta x=0.1$ is already a quite good approximation for the cellular dynamics in continuous limit.

\begin{figure}[thp]
\begin{center}
\includegraphics[width=80mm]{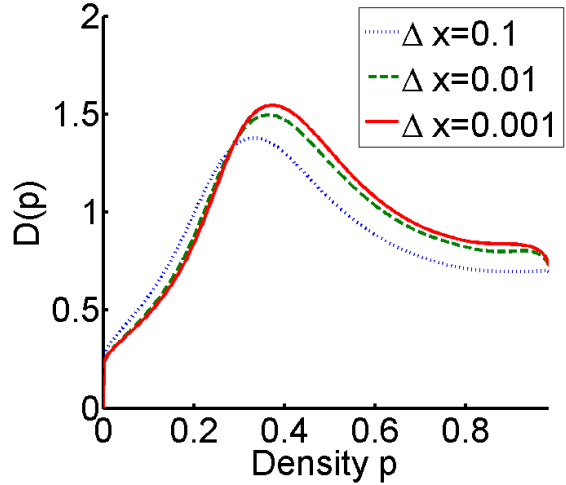}
\end{center}
\caption{Diffusion coefficient $D(p)$  generated by BM analysis at
$t_D=500$ for the same simulation as in Figure  \ref{densitydeltax}.
} \label{diffusiondeltax}
\end{figure}

The diffusion curves obtained from BM analysis are more sensitive to the change of $\Delta x$ compare with the sensitivity of $p(x)$ at Figure  \ref{densitydeltax}.
Figure \ref{diffusiondeltax} shows the diffusion curves from BM analysis obtained at $t_D=500$ from the same MSM simulations as in Figure  \ref{densitydeltax}.
It is seen that $D(p)$ is relatively well converges for $\Delta x \lesssim 0.01$. These changes in $D(p)$ do not change the efficiency of BM analysis for the prediction of density dynamics as in Figure \ref{diffusionCurves}.
For all values of $\Delta x=0.1$ the corresponding diffusion curves from BM analysis work well for the density dynamics. The effect of finite $\Delta x$ is only to modify
the   diffusion through additional $\Delta x$-dependent fluctuations of the reversal time $T$. The cause of such modification  is discussed  in Section
\ref{section:MicroscopicStochasticModelcellmotion}.

\begin{figure}[thp]
\begin{center}
\includegraphics[width=80mm]{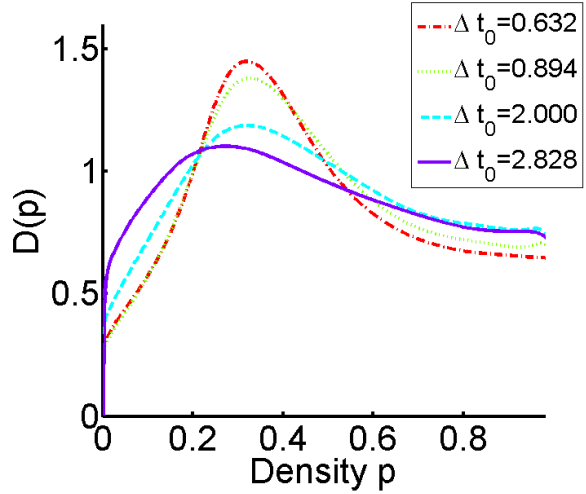}
\end{center}
\caption{Diffusion curves $D(p)$  for different values of variance $\Delta T_0$ of the reversal time $T$ fluctuations  (see the equation (\ref{Trandom}) and after it for definitions)
 obtained from BM analysis of ensemble-averaged MSM simulations. } \label{diffusionWNoise}
\end{figure}

\begin{figure}[thp]
\begin{center}
\includegraphics[width=80mm]{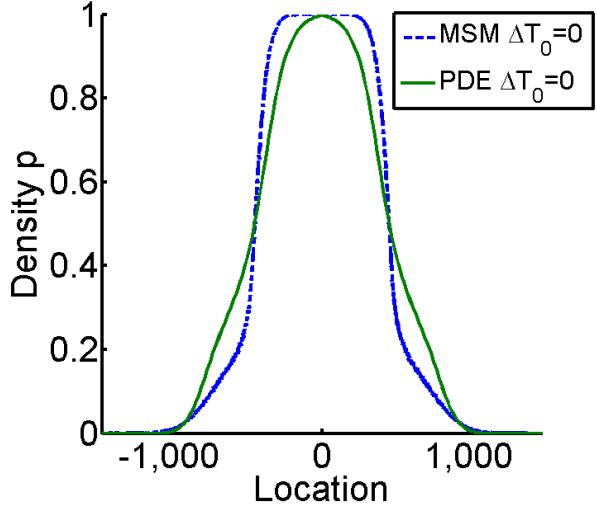}
\end{center}
\caption{Density curves for MSM simulations with $\Delta T_0=0$ and attempted PDE fit using BM analysis at $T=30,000$  and $T_D=500$. Initial condition had a form of top-hat distribution of width 1000.
} \label{BMPDENoNoise}
\end{figure}

We now address the effect of different values of $\Delta T_0$.
 In Myxobactera colony, the fluctuations of $T$
are possible with probability density
function for $T$ sharply peaked near average reversal period
$T$ \cite{Welch01}. We study the role of noise in the
reversal period by changing the value of  $\Delta T_0$.
For each $\Delta T_0$ we determine the nonlinear diffusion coefficient
 $D(p)$ using
BM analysis \cite{BoltzMat} of ensemble averaged MSM simulations. We found that $D(p)$ changes with $\Delta T_0$ as shown in Figure \ref{diffusionWNoise}.
Then we compared the density dynamics from MSM simulations and the nonlinear diffusion equation
(\ref{nonlineardiffusion1}).
For finite values of noise (typically for $\Delta T_0/T_0\gtrsim 0.1$) the agreement between  (\ref{nonlineardiffusion1}) and MSM simulations is very good, similar to shown in Figure  \ref{diffusionCurves}c.
 When no noise is added ($\Delta T_0=0$), we found that (\ref{nonlineardiffusion1}) is not a good approximation of the density dynamics, i.e.
 MSM density no longer follows (\ref{nonlineardiffusion1}) with  $D(p)$ from BM analysis as shown in Figure \ref{BMPDENoNoise}.

We conclude that the finite noise appears necessary for the nonlinear diffusion approximation to work.
Also note that that finite value of $\Delta x$ creates the effective noise in $T$ as discussed above. It means that generally effects of finite $\Delta T_0$ and finite $\Delta x$ add up. To distinguish these
effects one can reduce $\Delta x.$

\section{Analytical approximations of the pairwise and total collision times }
\label{section:Analyticalapproximation}


In this section an analytical approximation of the pairwise
collision time  and semi-analytical fit to the total collision times
are derived.   We mostly focus on a limit of intermediate cellular
density when $p>p_0$ but $p$ is not very close to one so that many
collisions between cells are still pairwise and they do not result
in larger clusters. Assume that a cell experiences on average jam(s)
of total duration $\tau_1$ from the left and $\tau_2$ from the right
during each reversal period $2T$. We include both pairwise jams and
indirect jams into definition of $\tau$ (see Section
\ref{section:elementarycollisions} for detailed definitions of
jams).  In such a case a shift of the center of mass of a given cell
per period $2T$ is $(\tau_1-\tau_2)v$.  The average collision time
$\tau$ in a given direction (left or right) must be a slow function
of $x$ (i.e. $\tau_1\simeq\tau_2\equiv \tau$) to avoid large
microscopic gradients. Typically $\tau$ can be viewed as the average
(ensemble or time average) over many collisions (jamming events) for
each given cell. It is necessary to stress that $\tau$ in this
section is the (total) jam time (from both direct and indirect jams)
per period $T.$ This quantity is different from $\tau_{cluster}$
from Section \ref{section:elementarycollisions} because $\tau$ never
exceeds $T$ by its definition.

Although jam times can  fluctuate strongly from collision to
collision  (as seen in Figure \ref{jam}),  after averaging over
several collisions $\tau$ becomes slow function of $x$ and $t$.
We also neglect for now the influence of the fluctuations of the reversal time, i.e. we assume that all reversal phases are constant. 
Below in this Section we also separately discuss the effect of these fluctuations. 
Taking into account finite value of $\tau$ we estimate the local
cellular density as $p=L/\langle L_{dist}\rangle$, where the average
distance between neighboring cells is $L_{dist}=L+v |\tilde
\phi_1-\tilde \phi_2|-v\tau$ and $\langle \ldots \rangle$ means
statistical averaging over the uniformly distributed phases $\tilde
\phi_1$ and $\tilde \phi_2$.  This expression is, however, only true
for $|\tilde \phi_1-\tilde \phi_2|\ge \tau$ because distance between
centers of mass of two neighboring cells is $\ge L$. For pairs of
cells with smaller difference in phases  $|\tilde \phi_1-\tilde
\phi_2|\le \tau$ we have to take into account simultaneous
collisions (jams) of three and more cells. During each triple
collision two neighboring cells have pairwise jams and  third one
has an indirect jam.  If $\tau$ is small $|\tilde \phi_1-\tilde
\phi_2|\le \tau$ then  cells 1 and 2 move most of the time parallel
to each other, either attached to each other or separated by a
typical distance $2v |\tilde \phi_1-\tilde \phi_2|$. After reversing
direction cell always alternates between these two possibilities. A
distance between average positions of the centers of mass of these
two cells is $\sim v|\tilde \phi_1-\tilde \phi_2|$. Cells 1 and 2
move almost all the time together separated by that average small distance
between them. After colliding with another (third) cell on the left
(referred to as cell 0) or with a cell on the right (referred to as
cell 3) they quickly form 3-cell cluster. Assume that lifetime of
each such cluster is about $\tau$. Then pairwise jam time for cells
0 (with cell 1) or cell 3 (with cell 2) is $\simeq \tau$. For cell 1
and 2 each collision is either a pairwise jam with the jam time
$\simeq \tau$ or a cluster jam with the jam time $\simeq \tau$. So
the average jam time is $\simeq \tau$ in both cases. The distance
between average positions of cells 0 and 1 (or between cell 2 and 3)
is $\simeq L+v(|\tilde \phi_1-\tilde \phi_2|-\tau)$.  Based on that
we obtain the following approximate expression for the average
distance between two neighboring cells combining contributions from
$ |\tilde \phi_1-\tilde \phi_2|\ge\tau$ and $|\tilde \phi_1-\tilde
\phi_2|\le \tau$:
\begin{eqnarray}\label{Laverage}
\langle L_{dist}\rangle=T^{-1}\left [\int^T_\tau (L+v[\phi-\tau])d \phi + \int^\tau_0[L+ v O(\phi)] d \phi\right ] =L +v\left (\frac{T}{2}-\tau\right )+v\tau O\left (\frac{\tau}{T}\right),
\end{eqnarray}
where we included the contribution of the average distance $\sim
v|\tilde \phi_1-\tilde \phi_2|$ between cells 1 and 2  for $|\tilde
\phi_1-\tilde \phi_2|\le \tau$ into $O(\phi)$ term. Here and below
by $O(x)$ we mean $O(x)=c_1 x+ c_2x^2+c_3x^3+\ldots$ with constants
$c_1, \, c_2, \ldots$ generally $\sim 1.$ Terms $\propto v\tau^2/T,$
$v\tau^3/T^2,$  $v\tau^4/T^3,$ etc. in  (\ref{Laverage}) result from
the 3-cell, 4-cell, 5-cell, etc. cluster contributions,
respectively. To establish scaling associated with the number of
cells in a cluster we note that probability to have n-cell cluster
is roughly proportional to the probability $P_{n-2}$ of $n-2$
neighboring cells simultaneously having small differences in phases
$|\phi_i-\phi_{i+1}|\lesssim\tau,$ $i=1, \, 2. \ldots, n-2$. Here
$P_{n-2}\propto (\tau/T)^{n-2}$ because phases are statistically
independent. n-cell cluster is formed by these $n-2$ cells together
with two surrounding cells involved in pairwise jams with the
average time $\tau$. Similar to the case of 3-cell cluster, the
$n-2$ cells inside of a cluster have average jam time $\tau$
dominated by the indirect jams. So the resulting contribution to the
$\langle L_{dist}\rangle$ is $\sim v\tau P_{n-2}$.   Of course for
densely packed clusters such approximation is oversimplifies but the
general form of  $O(x)$ remains the same. These qualitative
arguments do not affect the quantitative calculations described
below and yield qualitative understanding of the MSM dynamics.

The equation (\ref{Laverage}) results in the following relation
between cellular density and the collision time
\begin{eqnarray}\label{pan}
p_{an}=\frac{L}{\langle L_{dist}\rangle}=\frac{L}{L+vT/2-v\tau+v\tau O(\tau/{T})}.
\end{eqnarray}
After solving the equation (\ref{pan}) for $\tau$ we obtain the
analytical approximation for the average collision time
\begin{eqnarray}\label{tauantot}
\tau(p)=\left [\frac{T}{2}-\frac{L}{v \, p}+\frac{L}{v}+\tau O(\tau/{T})\right ]\Theta(p-p_0),
\end{eqnarray}
where $p_0$ is given by (\ref{p0def}), $\Theta(y)$ is the Heaviside step function ($\Theta(y)=1$ for
$y> 0$ and $\Theta(y)=0$ for $y<0$) and the factor $\Theta(p-p_0)$ results from the condition that $\tau\ge 0$ (recall that it is shown in Section  \ref{section:elementarycollisions} that jams are absent for $p<p_0$
if we neglect the fluctuations of the reversal time).

\begin{figure}
\begin{center}
\subfloat[]{\includegraphics[width=80mm]{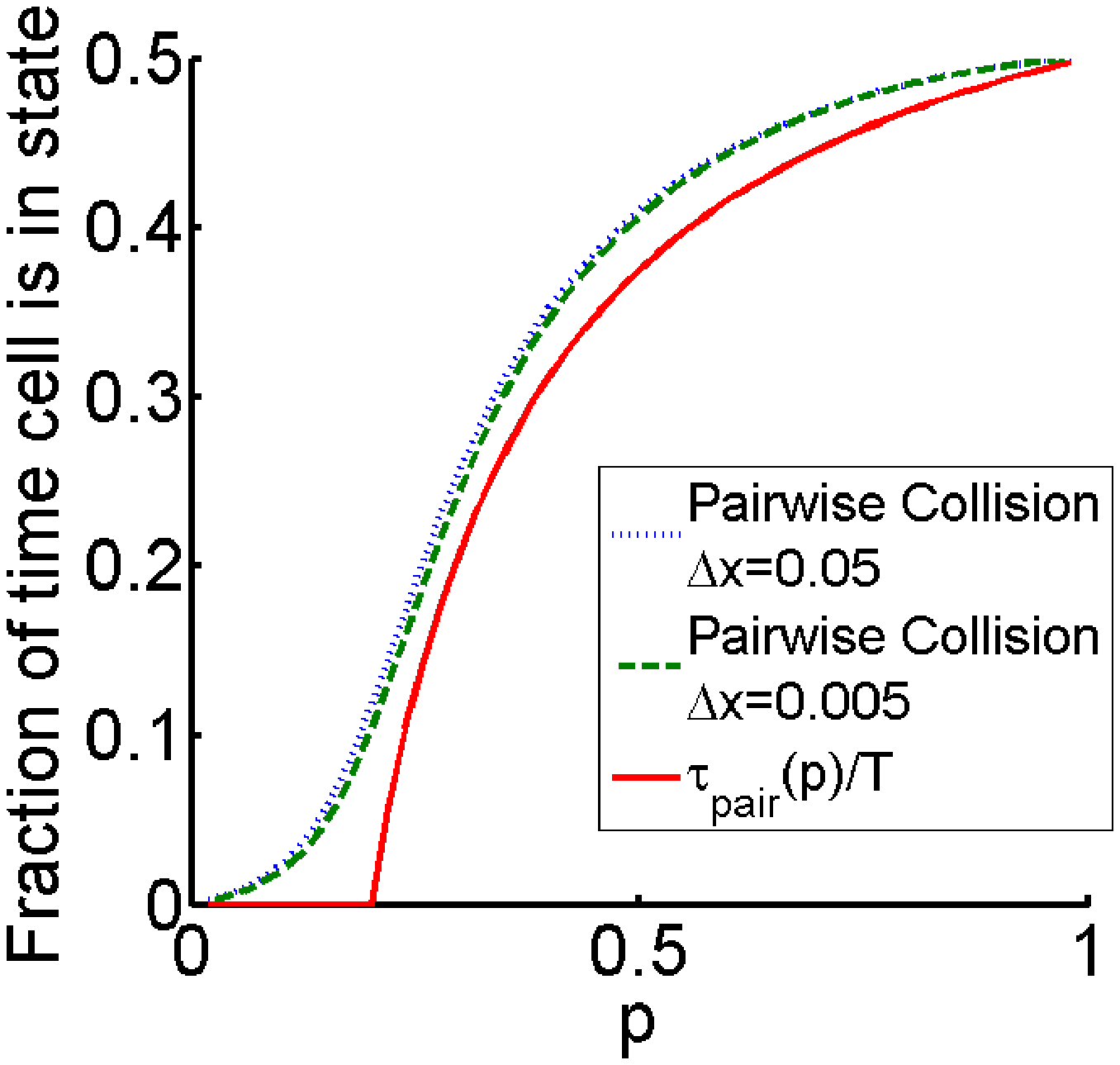}}
 \subfloat[]{\includegraphics[width=80mm]{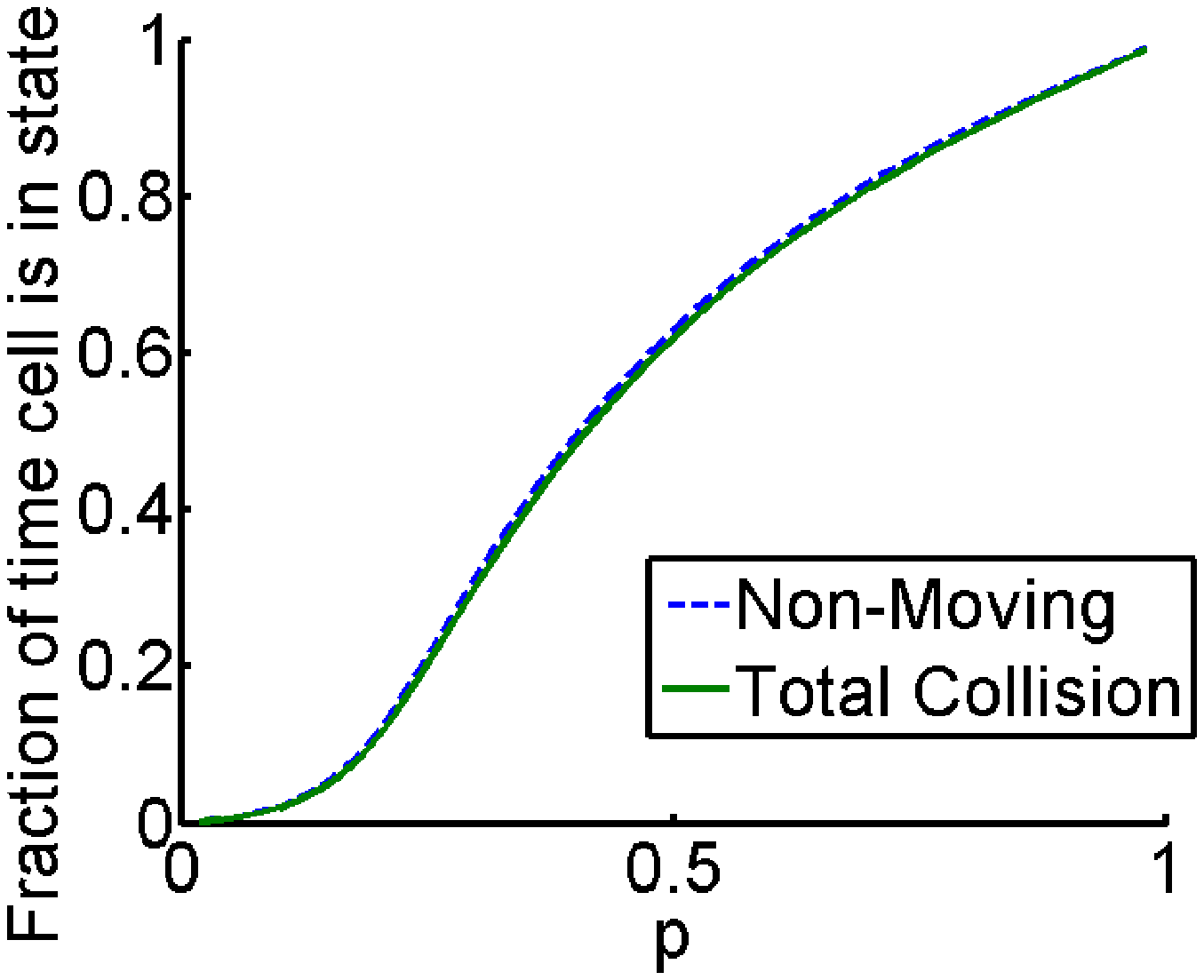}}
\end{center}
\caption{(a) Dependence of $\tau_{pair}(p)/T$ from MSM simulations with lattice size $\Delta x=0.05$ (dash-dotted line) and $\Delta x=0.005$ (dashed line) in comparison with the analytical expression (\ref{tauan}) for $\tau_{pair}/T$ (solid line) where $T=8$.  (b) Relative time cells spend non-moving per period $T=8$ (dashed line) vs. relative total jam time $\tau/T$ (solid line) with $\Delta x=0.005$. $\tau$ includes both indirect and pairwise collisions. Decrease of $\triangle x $ results in better match between each pairs of curves.}
\label{fig:collisiontime}
\end{figure}

Neglecting $\tau O(\tau/{T})$ in (\ref{tauantot}) means that we take
into account pairwise jams only and neglect indirect jams resulting
in the average pairwise collision time $\tau_{pair}$:
\begin{eqnarray}\label{tauan}
\tau_{pair}(p)=\left [\frac{T}{2}-\frac{L}{v \, p}+\frac{L}{v}\right ]\Theta(p-p_0).
\end{eqnarray}
Figure \ref{fig:collisiontime}a
compares $\tau_{pair}(p)/T$ simulations obtained using MSM with
simulations from (\ref{tauan}). MSM simulations were performed with
the periodic boundary conditions at the spatial interval of length
$1000$ and initial random placement of $N$ cells (avoiding
configurations forbidden by excluded volume principle). We used
$\triangle x=0.05$ and $\triangle x=0.005$. $N$ was chosen for each
simulation to match given $p$ (i.e. $N=1000 \, p$). All types of
collision times were calculated by running simulations till the
final simulation time $t_{final}=10^6$. We also assumed ergodicity
and recorded collisions of all cells during each such simulation.
Convergence was tested by comparing the results from a subset of
densities to results obtained with $t_{final}=10^7$ and a good match
was demonstrated. Ergodicity was also tested by comparing the
collision time results from several different stochastic
realizations with $t_{final}=10^6$ and a very good match was shown
for tested density values.

Figure \ref{fig:collisiontime}a shows that MSM simulations and
(\ref{tauan}) are in a reasonably good agreement for $p>p_0$. For $p<p_0$ we generally need to modify (\ref{tauan}) to include the fluctuations of the reversal time. Such modification is outside the scope of this paper.
We however can immediately estimate the order of such modification. E.g., for $p=p_0$ cells do not collide without fluctuations of $T$ but they often come  to zero distance between them and move in parallel as
explained in Section \ref{section:elementarycollisions}. It means that if we include fluctuations of $T$ then the typical pairwise collision time will be $\sim\Delta T_0$. Respectively $\tau_{pair}(p_0)$ takes a value
$\tau_{pair}(p_0)\sim \Delta T_0$ (in contrast with $\tau_{pair}(p_0)=0$ in (\ref{tauan}))  in good agreement with Figure \ref{fig:collisiontime}a ($\Delta T_0/T=0.09$ for the parameters of MSM
of Figure  \ref{fig:collisiontime}a while  $\tau_{pair}(p_0)/T\simeq 0.1$ for the dashed curve of Figure  \ref{fig:collisiontime}a). For $p<p_0$ such modification is decreased because for smaller density cells collide
only due the random walk from the fluctuations of $T$. Smaller the density, more time it takes for random walk to ensure collisions. It results in the decay of $\tau_{pair}(p)$ to zero as $p\to 0.$
For $p>p_0$ the fluctuations of $T$ results in more efficient exploration of the space by cells which increase  $\tau_{pair}$ compare with (\ref{tauan}).
This explains the difference between solid curve and dashed curve in Figure \ref{fig:collisiontime}a.

We also performed simulations  with decreasing
values of $\triangle x$ to demonstrate that $\triangle x=0.05$ is already small enough to give a good approximation of $\tau_{pair}(p)$ compare with the continuous limit $\triangle x\to 0$.
Figure\ref{fig:collisiontime}a  shows that with dashed and dotted curves corresponding to $\tau_{pair}(p)$  for    $\triangle x=0.005$ and  $\triangle x=0.05$, respectively.

The same MSM simulations were used to calculate the total collision
time per period $T$. In simulations we distinguish two types of the
total collision time. The first type is the total collision time
$\tau$ itself (total jam time per period $T$) which includes both
pairwise jams and indirect jams. The second type is the average time
(per period $T$) cells spend without movement which includes
pairwise jams, indirect jams and jams due to finite value of
$\triangle x$. The third contribution occurs when two cells are
attached to each other and move in the same direction. If a cell
which moves behind second one, is chosen by the MSM algorithm then
its movement is prevented by the second cell. This artificial effect
is due to discretization and finite value of $\triangle x$. It
disappears for $\triangle x\to 0$ so that both types of the total
collision time are the same in that limit. Figure
\ref{fig:collisiontime}b shows  the time cells spend without
movement per period $T$ versus $\tau$.   It demonstrates that these
total collision times (normalized to $T$) are very close to each
other for $\triangle x\to 0.005$.
\begin{figure}
\begin{center}
\includegraphics [width=80mm]{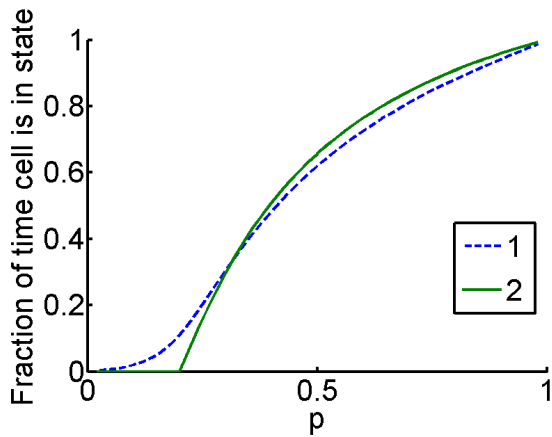}
\end{center}
\caption{Plots of the total collision time per period $\tau/T$ from MSM simulations
(dashed line 1) and from the analytic approximation
of $\tau_{approx}$ $(\ref{tauanapprox})$ (solid line 2).
The lattice size is $\triangle x =0.005$ and $T=8$.}
\label{fig:clustercollisiontime}
\end{figure}

In contrast to  the pairwise collision time used in (\ref{tauan}),
the equation (\ref{tauantot}) includes an extra term $\tau
O(\tau/{T})$ which  corresponds to the total  jam time $\tau$ per
period $T$. Dashed line in Figure \ref{fig:clustercollisiontime}
shows $\tau(p)$ dependence in MSM simulations with the noise in $T$ (with the standard $\Delta T_0=0.9$).
 We now approximate
the term $\tau O(\tau/{T})$ in the equation (\ref{tauantot}) in the
simplest possible way neglecting noise in $T$ as $O(x)=c_1x$ with $c_1=2$ which yields
\begin{eqnarray}\label{tauanapprox}
\tau_{approx}(p)=\tau_{pair}(p)\left [1+2\frac{\tau_{pair}(p)}{T} \right ]\Theta(p-p_0),
\end{eqnarray}
where $\tau_{pair}(p)$ is given by  (\ref{tauan}). $\Theta$-function  reflects the neglect of noise in $T$ similar to (\ref{tauan}). We choose $c_1=2$
here to ensure correct asymptotic value $\tau_{approx}(1)=T$  for
$p=1$ because all cells are jammed all the time in that case. Figure
\ref{fig:clustercollisiontime} shows a reasonably good fit between
(\ref{tauanapprox}) (solid line 2) and the total collision time per
period $\tau/T$ from MSM simulations with $\Delta T_0=0.9$  (dashed line 1). It suggests
that if we add an effect of noise in (\ref{tauanapprox}) then it might become very good fit.  Exact analytical theory is  needed in order to verify
this hypophysis which is quite a challenging problem and which is
outside the scope of this article.

\section{Conclusions and Discussion}
\label{section:ConclusionDiscussion}

In this paper, a connection was established between stochastic 1D
model of microscopic motion of the system of regularly reversing
self-propelled rod-shaped cells and a nonlinear diffusion equation
describing macroscopic behavior of this system. Macroscopically
(ensemble-vise) averaged stochastic dynamics was shown to be in a
very good agreement with the numerical solutions of the nonlinear
diffusion equation (\ref{nonlineardiffusion1}), where the diffusion
coefficient was obtained using BM analysis. Critical density $p_0$
was found such that for $p<p_0$ the
cellular diffusion is dominated by the diffusion (random walk) of individual cells while for $p>p_0$ the diffusion is dominated by the collisions between cells.
$p_0$ was determined (\ref{p0def}) from the
condition that cells do not jam with each other in the no noise limit. We  found that the role of noise in the reversal period is crucial.
Without noise, BM analysis cannot reproduce the MSM dynamics which means that nonlinear diffusion
is not a good approximation for the MSM dynamics. However, even  relatively small level ($\Delta T_0/ T\simeq 0.1$) of such noise produces
excellent agreement between BM based nonlinear diffusion and MSM simulations. The primary role of such small noise is to ensure randomization of collisions between different cells in the system
at large  in comparison with the reversing period $T$ times.

An analytical approximation of the pairwise collision time
$\tau_{pair}$ (\ref{tauan})  and semi-analytical fit for the total
jam time per reversal period $\tau_{approx}(p)$ (\ref{tauanapprox})
have been also obtained. Frequent collisions for $p>p_0$ are
responsible for the nonlinear diffusion of  the cellular density.
For $p<p_0$ cells tend to spread out so their collisions are possible only if we take into account
the fluctuations of the reversal time. Without such fluctuations there are no collisions
and no cellular transport is possible because cells experience
periodic motion in space and time. There still remains quite a
challenging problem of developing  a full statistical theory of 1D
self-propelled rod dynamics with reversals which would be applicable
for all densities. Such theory would require a detailed description
of formation and interaction of large cellular clusters.

It was also shown that nonlinear diffusion coefficient $D(p)$ used
to describe the macroscopic process, changes depending on the
reversal period. Small and large reversal periods yield diffusion
coefficients that favor high and low density diffusion respectively
as is shown in Figure  \ref{PDEDensityReversals}. Since dynamics of
the system is determined by the dimensionless parameters  $vT/L$
(the ratio of distance traveled by cells between reversals  and the
cell length) and $p$, increase of the speed at which cells move is
equivalent to the increase of the reversal period. Thus, cell
populations with small $T$ are able to spread out effectively at
high densities while large $T$ promotes cell population swarming at
smaller densities.

An interesting problem to be studied in future work is to determine
the optimal choice of reversal time $T$ maximizing the swarming rate
of Myxobacteria colony using nonlinear diffusion equation, and
compare it with the one  obtained in \cite{Wu09} using stochastic
model.

\appendix
\section{Boltzmann-Matano Analysis}
\label{section:BoltzmannMatanoAnalysis}

In this appendix we review Boltzmann-Matano (BM) Analysis (see
\cite{BoltzMat} for details) for the readers convenience. Assume
that the process we are studying can be modeled using the nonlinear
diffusion equation (\ref{nonlineardiffusion1}) with some unknown
nonlinear diffusion coefficient $D(p)$.

BM analysis allow to recover $D(p)$ from the 1D dynamics of the cellular density $p$ with the stepwise initial condition
\begin{equation}\label{pinistep}
       p(x,0)=\begin{cases} p_L,\quad & \text{if } \ x<x_M\\
                                p_R &  \text{if }\ x>x_M
                        \end{cases}.
\end{equation}
at infinite 1D domain. Here we assume that $p_L>p_R$.

Special property of the stepwise initial condition is that it does not have any spatial scale (spatial size of system is infinite and spatial scale of jump at $x=x_M$ is zero.
Then the only possible solution has a self-similar form $\rho(\zeta)$ which was found by Boltzmann in 1894. Here
\begin{align} \label{zetaEqndef}
\zeta=(x-x_M)/t^{1/2}
\end{align}
which is motivated by a self-similar solution of a heat equation (for $D=const$). $x_M$ is a reference point also known as the Matanos interface. Assuming that $p(\zeta)$ does not depend on $t$ explicitly, we obtain that $\displaystyle \frac{\partial}{\partial t}p(\zeta)=-\frac{1}{2}\frac{\zeta}{t}\frac{\partial}{\partial {\zeta}}p(\zeta)$
and $\displaystyle \frac{\partial}{\partial x}p(\zeta)=\frac{1}{t^{1/2}}\frac{\partial}{\partial {\zeta}}p(\zeta)$ which allows to reduce  (\ref{nonlineardiffusion1}) to
\begin{align} \label{zetaEqn}
-\frac{\zeta}{2}\frac{\partial}{\partial {\zeta}}p=\frac{\partial}{\partial {\zeta}}\Big [ D(p)\frac{\partial}{\partial {\zeta}}p\Big ].
\end{align}
Since the solutions to a non-linear diffusion equation with
stepwise initial conditions are monotonic, it follows that for any
given fixed time the function $p(x)$ is invertible with respect to $x$.
Below we use the notation $x(p)$ for the inverse of $p(x)$.
Integrating both sides of (\ref{zetaEqn}) with respect to $\zeta$
yields
$$\displaystyle -\frac{1}{2t^{1/2}}\int_{p_L} ^{p} (x(p)-x_M) dp=D(p)p_{\zeta}$$
\noindent where the left hand side follows from
$$\displaystyle\int_{-\infty}^{\zeta} \zeta \frac{\partial p}{\partial {\zeta}}  d\zeta= \int_{p_L} ^{p} \zeta(p) dp=\frac{1}{t^{1/2}}\int_{p_L} ^{p} (x(p) -x_M) dp.$$
\noindent Since $\frac{\partial p}{\partial \zeta} =
t^{1/2}\frac{\partial p}{\partial x}$, the equation can be rewritten
as
$$\displaystyle D(p)=-\frac{1}{2t}\left [\frac{\partial p}{\partial x}\right ]^{-1}\int_{p_L} ^{p} (x(p)-x_M) dp,$$
\noindent which gives the Boltzmann description of the diffusion
equation. Now it is possible to calculate the appropriate value of
the interface, $x_M$, to ensure that the diffusion calculation is
consistent. Specifically, since mass diffuses from the left to the
right across the interface, there is a mass conservation equation
where the mass lost on the left of the interface should equal the
mass gained on the right of the interface,
$$\int_{-\infty}^{x_M} (p_L- p(x))dx = \int_{x_M}^{\infty} (p(x) - p_R)dx.$$
\noindent Again inverting $p(x)$, we can calculate the area under of
the integrals in terms of $x(p)$ to get the following equivalent
expression
$$\int_{p_M}^{p_L} (x(p)-x_M) dp = \int_{p_R} ^ {p_M} (x_M-x(p)) dp,$$
\noindent which simplifies to
$$\int_{p_L} ^{p_R} (x(p)-x_M) dp=0.$$
\noindent Mass conservation occurs precisely when
\begin{align} \label{xmdef}
x_M=\frac{\int_{p_L} ^{p_R} x(p) dp}{p_L-p_R},
\end{align}
which is the Matano's result to determine $x_M$ if it is unknown in advance.

In our simulations we know $x_M$ in advance so in fact we use Boltzmann analysis, but not BM analysis (except additional tests discussed in Appendix B).
Also in our simulations $p_L=p_{max}$ and $p_R=0.$

\section{Accuracy of Boltzmann-Matano Analysis}

 \begin{figure}
  \begin{center}
 \subfloat[]{\includegraphics[height=50mm]{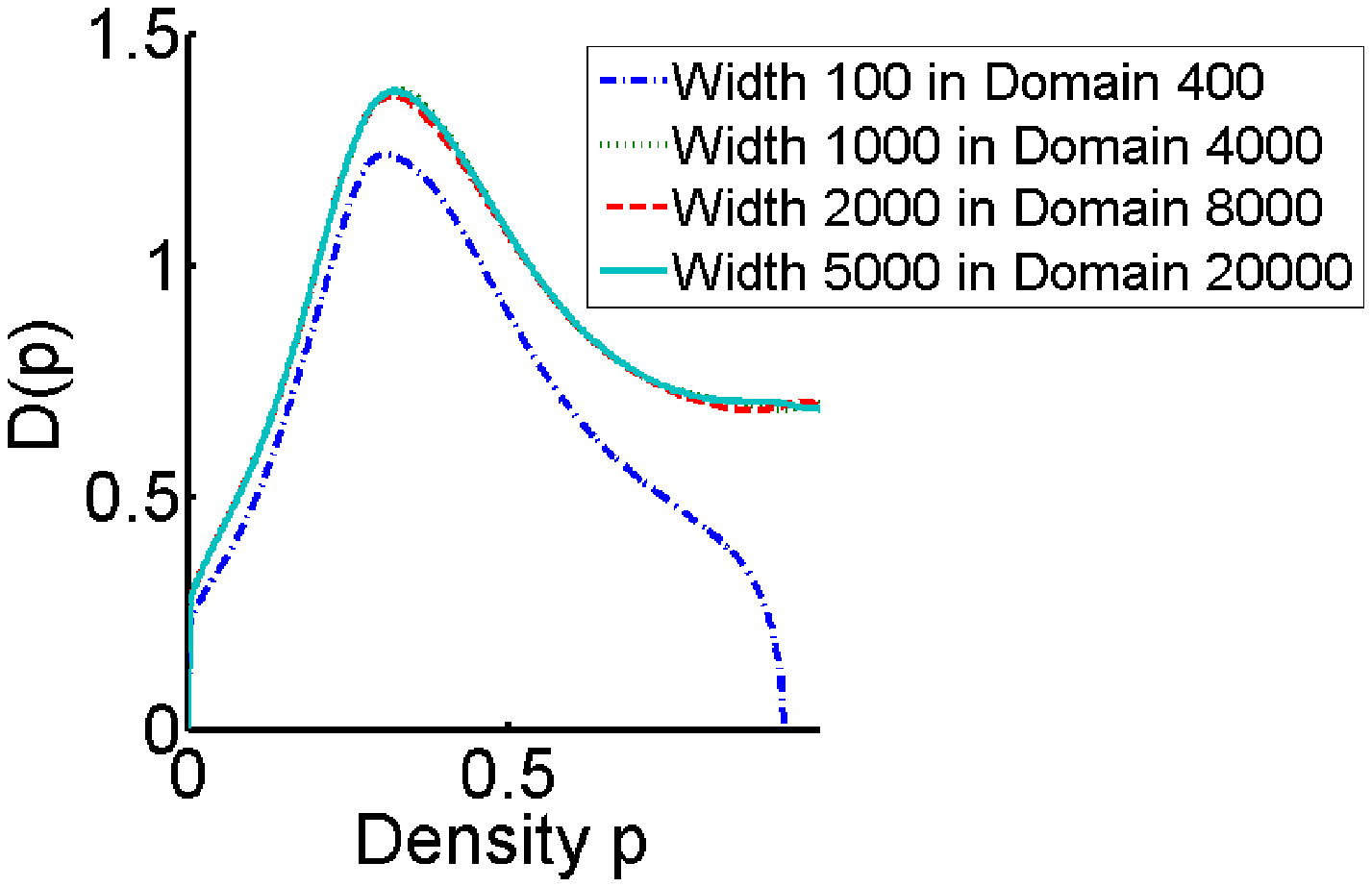}}\quad
 \subfloat[]{\includegraphics[height=50mm]{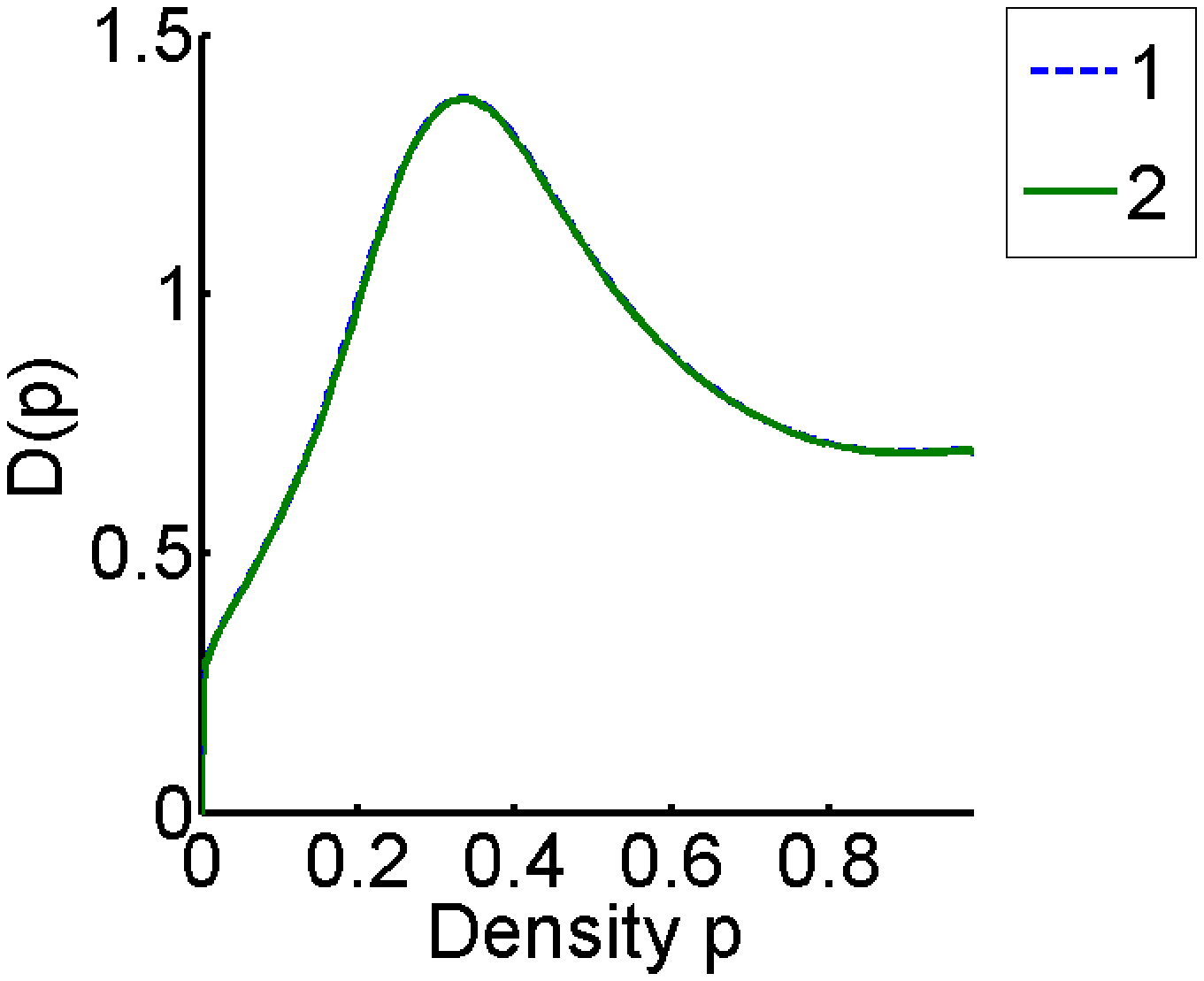}}\\
 \subfloat[]{\includegraphics[height=50mm]{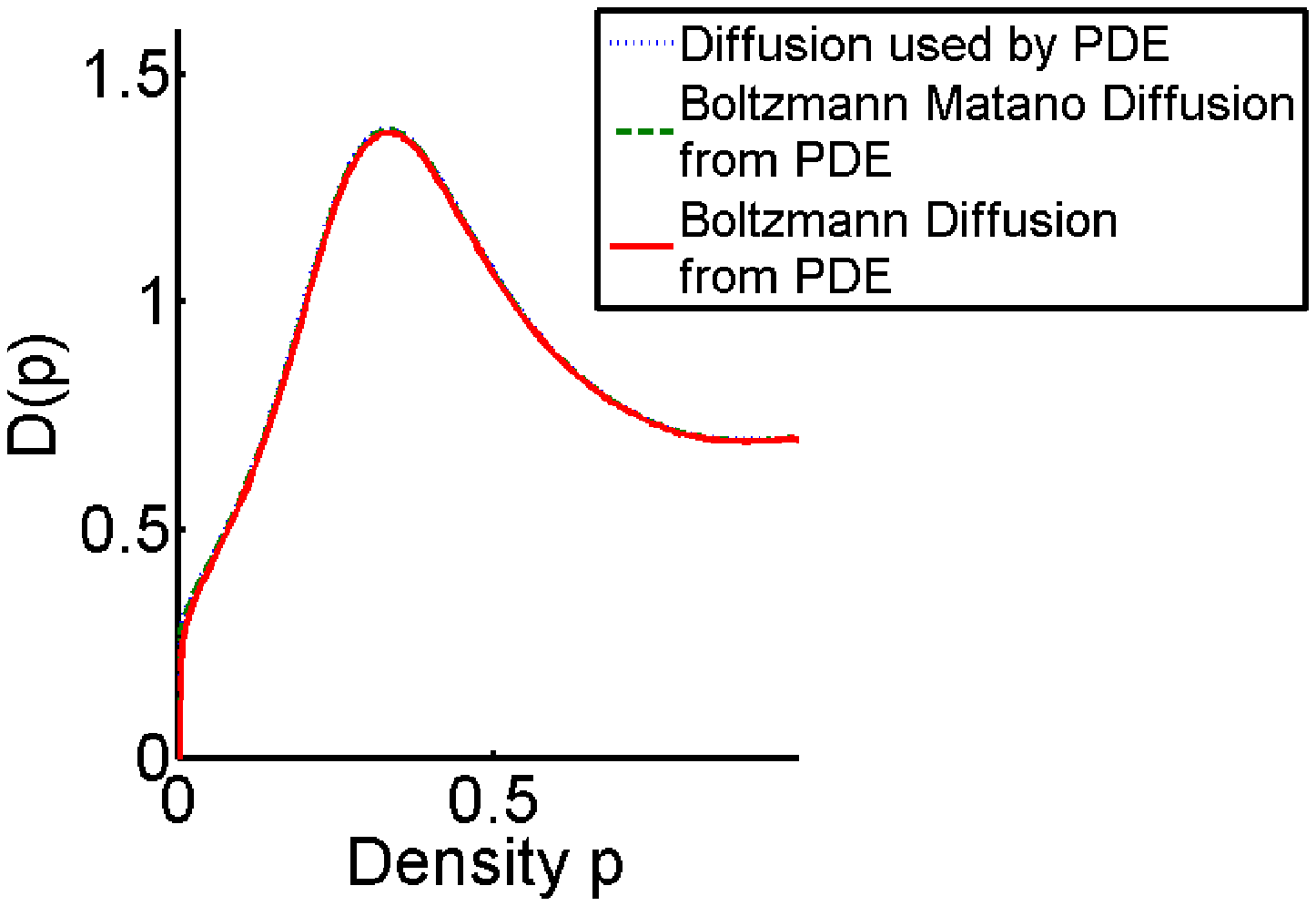}}
\end{center}
\caption{(a) The nonlinear diffusion coefficient $D(p)$ determined from BM analysis of MSM simulations
with different initial top-hat widths. The density profiles at $t_D=500$ were used for all curves. It is seen that curves for the widths 1000 and above are almost undistinguishable.
(b)  $D(p)$ obtained from MSM simulations (dashed line - curve 1) and PDE simulation (solid line - curve 2) for top-hat initial conditions of width $600$. Density profile at time $t_D=1,000$ is used for BM analysis.
 (c) Comparison of BM analysis with Boltzmann analysis from PDE density profile at $t_D=500$. Dashed line is $D(p)$ used to produce density profiles from PDE simulations.
 All curves at (b) and (c) are almost indistinguishable.}\label{BMTests}
\end{figure}

BM analysis, described in Appendix A, is defined on infinite spatial interval with step-wise initial conditions only. Assume now that we apply BM analysis for  top-hat initial conditions as described in Section \ref{section:MicroscopicStochasticModelcellmotion}.  In that case BM analysis is only approximate one because initial conditions include spatial scale $x_{width}$, which is the spatial width of top-hat. Self-similar solution of Appendix A does not agree with top-hat.  That solution is only approximately valid in the neighborhood of each of two steps of top-hat. Because of spatial symmetry it is enough to consider any of these two steps.
To estimate the accuracy of BM analysis in that case we note that if the density at $x=0$ (middle of top-hat) remains nearly constant then BM analysis is still applicable (except small unavoidable corrections because for any $t>0$ density is never exactly constant). Assuming that the diffusion coefficient $D(p)\sim 1$, we roughly estimate that the width of initial top-hat doubles with time when $D(p)t_0/ x_{width}^2\sim 1$ which gives $t_0\sim 10^6$ for  $x_{width}=1000$. For $t\ll t_0$ a change of density in the middle of top-hat is small in agreement with Figure \ref{diffusionCurves}.
A similar limitation of BM analysis is that the total spatial width of the simulation domain must  exceed the width of top-hat in several times to make sure that the cellular density remains low at boundaries as seen in   Figure \ref{diffusionCurves}.

As additional test of BM analysis we  varied the domain length and width
of the initial top-hat distribution calculating diffusion coefficient by BM analysis from MSM simulations (Figure \ref{BMTests}a). We observed that small top-hat width $\sim 100$ is not
enough for applicability of BM analysis(dash-dotted curve in Figure \ref{BMTests}a) while top-hat widths $\gtrsim 1000$ total domain lengths  $\gtrsim 4000$  are far enough for such applicability.
Figure  \ref{BMTests}b compares $D(p)$ obtained from PDE simulation (solid line) and MSM simulations (dashed line) for top-hat initial conditions of width $600$. Difference between these curves is almost
indistinguishable.  This indicates that our statistical ensemble in MSM simulations is large enough to avoid influence of noise in the data on the diffusion curve.  We also tested MSM data with
and without the Gaussian filter and obtained the same diffusion curves.
Larger widths were also tested and proven to
 match very well, but the results are not displayed here. From
these observations, we can conclude that the generated diffusion
curves are independent of the width of the top hat used if the top
hat is sufficiently long enough in such a way that the center and
boundaries have constant density.

We would like to point out to avoid confusion that we need BM analysis only to determine the diffusion curves at reasonably small times ($t \ll t_0$).
After that we run  PDE simulations with these diffusion curves for much longer time (when density is changing both at the middle of top-hat and at boundaries).
For these much larger times we also see very good agreement between MSM simulations and PDE simulations (see e.g. Figures \ref{8reversalperiod} and \ref{PDEDensityReversals}).

As discussed in Section  \ref{section:MacroscopicNonlinearDiffusionModelvsMSMsimulations}, another limitation of BM analysis is the loss of numerical precision near $p(x)=const $ because  BM analysis
requires calculating $(dp(x)/dx)^{-1}$. Many Figures (\ref{diffusionCurves} and \ref{BMTests}) have jumps of $D(p)$ near $p=1$ which is due to such loss of numerical precision which can be
fixed by the polynomial extrapolation. That is however not necessary because these jumps do not change results of PDE simulations in any significant way.


We also tested BM analysis vs. Boltzmann analysis as shown in Figure \ref{BMTests}c. Although we know $x_M$ from top-hat initial conditions, but for finite width of top-hat we can ask if allowing $x_M$ to be
located not exactly at the step of top-hat could improve the accuracy of BM analysis to determine  $D(p).$ In that sense we can consider $x_M$ as additional fitting parameter
to accomodate finiteness of top-had width. Figure \ref{BMTests}c compares diffusion curves obtained from BM and Boltzmann analysis vs. exact diffusion curve. We see that difference
in accuracy between BM and Boltzmann analysis is very small.  It appears the advantage of using BM analysis vs. Bolztmann analysis is not significant in our case.

\section*{Acknowledgments}

Work of P.L. was supported by NSF grant DMS 0719895 and UNM RAC grant. R.G. was supported by
the University of Notre Dame's CAM Fellowship and partially supported by NSF grant DMS 0931642. M.A. was partially supported by the NSF grants DMS 0931642 and BCS 0826958.


\end{document}